\documentclass[10pt,journal,compsoc]{IEEEtran}

\usepackage{tabularx}
    \newcolumntype{L}{>{\raggedright\arraybackslash}X}
\usepackage{comment}
\usepackage{amsmath}
\usepackage{amssymb}
\usepackage{pifont}
\usepackage{gensymb}
\usepackage{graphicx}
\usepackage{svg}
\usepackage{epstopdf} 
\usepackage{color}
\usepackage[bookmarks=false,hidelinks]{hyperref}
\usepackage{enumerate}
\usepackage{multirow}
\usepackage{cite}
\usepackage{multicol}
\usepackage{threeparttable}
\usepackage{enumerate}
\usepackage[ruled, lined]{algorithm2e}
\usepackage{subfig}
\usepackage{float}
\usepackage{mathtools}
\usepackage{array}
\newcolumntype{P}[1]{>{\centering\arraybackslash}p{#1}}
\newcolumntype{M}[1]{>{\centering\arraybackslash}m{#1}}
\newcommand{\cmark}{\ding{51}}
\newcommand{\xmark}{\ding{55}}
\usepackage[numbers]{natbib} 

\begin{document}

\title{GRAVITAS: Graphical Reticulated Attack Vectors for Internet-of-Things Aggregate Security}

\author{Jacob Brown, Tanujay Saha, and Niraj K. Jha~\IEEEmembership{Fellow,~IEEE}
\thanks{
This work was supported by NSF under Grant No.~CNS-1617628.  Jacob Brown, Tanujay Saha, and
Niraj K. Jha are with the Department of Electrical Engineering, Princeton University, Princeton,
NJ, 08544 USA, e-mail:\{jacob.brown,tsaha,jha\}@princeton.edu. IEEE Copyright Notice:
© 2021 IEEE. Personal use of this material is permitted. Permission from IEEE must be obtained for all other uses, in any current or future media, including reprinting/republishing this material for advertising or promotional purposes, creating new collective works, for resale or redistribution to servers or lists, or reuse of any copyrighted component of this work in other works. This paper has been accepted to be published in IEEE Transactions on Emerging Topics in Computing, 2021.}
}

\IEEEtitleabstractindextext{
\begin{abstract}
Internet-of-Things (IoT) and cyber-physical systems (CPSs) may consist of thousands of devices 
connected in a complex network topology. The diversity and complexity of these components present an 
enormous attack surface, allowing an adversary to exploit security vulnerabilities of different 
devices to execute a potent attack. Though significant efforts have been made to improve the security 
of individual devices in these systems, little attention has been paid to security at the aggregate 
level. In this article, we describe a comprehensive risk management system, called GRAVITAS, for IoT/CPS that 
can identify undiscovered attack vectors and optimize the placement of defenses within the system for optimal performance and cost. 
While existing risk management systems consider only known attacks, our model employs a machine learning 
approach to extrapolate undiscovered exploits, enabling us to identify attacks overlooked by 
manual penetration testing (pen-testing). The model is flexible enough to analyze practically any IoT/CPS and provide 
the system administrator with a concrete list of suggested defenses that can reduce system vulnerability at optimal cost. 
GRAVITAS can be employed by governments, companies, and system administrators to design secure IoT/CPS
at scale, providing a quantitative measure of security and efficiency in a world where IoT/CPS devices will soon be ubiquitous.
\end{abstract}

}

\maketitle
\IEEEdisplaynontitleabstractindextext
\IEEEpeerreviewmaketitle

\IEEEraisesectionheading{\section{Introduction}}
\label{Intro}
\IEEEPARstart{I}{nternet}-of-Things (IoT) refers to any system that comprises multiple Internet-connected devices that provide transmission or computational services as one networked entity 
\cite{mosenia2017comprehensive}. Cyber-physical systems (CPSs) employ sensor data to monitor the 
physical environment and create real-world change using actuators. These broad categories include 
systems ranging from a single Bluetooth-enabled smartwatch to a smart city  containing millions of 
devices. Many of these devices employ rudimentary operating systems and are energy-constrained, 
often lacking basic security features. IoT/CPS may also consist of a diverse 
set of devices and complex network topologies, presenting a large attack surface that provides 
multiple enticing opportunities for a cunning adversary. The unique IoT/CPS exploits require that organizations employ rigorous threat analysis and risk mitigation techniques to minimize the likelihood of a successful attack.

\par Organizations are projected to have spent \$742 billion USD on IoT devices in 2020 alone $-$ an 
amount which is expected to increase by 11.3\% annually through 2024 \cite{idc2020}. Over the next 
few years, we can expect to see IoT/CPS move beyond small-scale applications and become significantly 
more commonplace in healthcare, manufacturing, transportation, law 
enforcement, and energy distribution
\cite{mosenia2017comprehensive, bosche2018, akmandor2018smart, stojkoska2017review, zhang2014iot, 
yun2010research, al2015role, datta2016integrating}. The development of 5G communication infrastructure, 
autonomous vehicles, and hardware specifically designed for machine learning (ML) is also accelerating 
the adoption of large-scale IoT/CPS \cite{thierer2015}. 

\par However, many industry experts and leading political figures argue that the widespread adoption 
of IoT systems has the potential to engender ``catastrophic"  consequences \cite{stavridis2016, 
lewis2016, markey2020}. One ominous sign is the Mirai botnet attack of 2016, a distributed 
denial-of-service (DDoS) attack that briefly brought down large parts of the Internet on the U.S. 
East Coast \cite{margolis2017}. This attack was particularly notable because a single security
flaw $-$ unchanged default passwords $-$ resulted in significant technological and 
economic disruption. This catastrophic outcome highlights how a single malicious adversary can 
potentially compromise an entire IoT/CPS, if not the whole Internet \cite{margolis2017}. 

\par Cyber attacks like the Mirai botnet should serve as a warning: every IoT/CPS must be scrutinized 
for exploit pathways before deployment. The large attack surface of autonomous vehicle networks, smart 
cities, and other publicly-accessible IoT/CPS should draw particular scrutiny because a security 
breach of at least one connected device during the lifetime of the system is extremely likely, potentially allowing the adversary to wreak havoc on 
other parts of the system. It is not uncommon for multiple devices in an IoT system to have
various vulnerabilities. An attacker can utilize these vulnerabilities to launch multi-stage
attacks. In multi-stage attacks, a compromised device can be used as a stepping stone to attack other 
devices in the network. In order to prevent future Mirai-like attacks, engineers will need to take 
into account not just the security of individual devices, but of the system as a whole. 
\par
Moreover, securing IoT/CPS is challenging because of the limited resources available to their constituent devices. Such limitations often preclude the devices from employing intrusion detection mechanisms and executing complex cryptographic protocols. Although IoT-friendly lightweight protocols exist ~\cite{sehwag2016tv,mckay2016report}, it is still challenging to select a suitable combination of defenses to obtain optimal performance and security of the system. The high cost of adding defenses to large IoT/CPS is a frequent obstacle in deploying security features, making it critical for organizations to maximize risk reduction given a limited security budget.

\par Our model, called GRAVITAS, overcomes these challenges by combining the hardware,
software, and network stack vulnerabilities of a system into a single attack graph. This attack
graph, which includes undiscovered vulnerabilities and the connections
between them as predicted by the SHARKS ML model \cite{saha2020}, contains attack vectors that
are passed over by risk management tools that employ only known vulnerabilities. These attack vectors are then assigned risk scores according to a probabilistic method that models the interaction between attack impacts and the graph's vulnerabilities. Using these quantitative scores as a foundation for measuring risk, GRAVITAS suggests defenses to the system using an optimization process that lowers the risk score at minimum cost. With an IoT/CPS design and threat model as input, and a list of the most cost-effective defenses as output, GRAVITAS presents a security model that allows the system administrator to discover new attack vectors and proactively design secure IoT/CPS both pre- and post-deployment.

\par The novelty of the proposed methodology lies in:

\begin{itemize}
    \item An automated IoT/CPS-specific exploit discovery tool that includes potential vulnerabilities and attack vectors that have yet to be discovered.
    \item A novel exploit scoring system that uses the topology of vulnerabilities in the attack graph to gauge risk at both the exploit and device levels.
    \item The ability to suggest the most effective defenses at an optimal cost via an optimization algorithm tailored to the threat model.
    
\end{itemize}

\par The article is organized as follows. Section~\ref{section:Related Work} describes related work 
that informs this article. Section~\ref{section:Background} presents background material on our threat model and the
ML-generated attack graph that serves as the foundation for GRAVITAS, as well as a primer on attack 
graphs and the Common Vulnerability Scoring System (CVSS). 
Section~\ref{section:Motivation} provides a brief 
description of how GRAVITAS provides novel capabilities to IoT/CPS administrators. 
Section~\ref{section::Methodology} gives details of our methodology, including the inputs and 
model outputs. Section~\ref{section:example_application} provides a practical example of the model's 
functionality. Section \ref{section:discussion} includes a discussion and ideas for future work. 
Section \ref{section:conclusion} concludes the article.

\section{Related Work}
\label{section:Related Work}

Most IoT-related security research to date has focused on remediating device-specific or
application-specific security vulnerabilities. Over the last decade, researchers have
discovered eavesdropping on implanted medical devices, ``outage"  attacks on IoT systems in
nuclear power plants, tampering with smart home devices, identity theft using corrupted RFID
tags, and poisoning ML models by changing sensor data, among many others \cite{nia2016, matrosov2011,
hernandez2014, juels2003, huang2011}. This research also occurs in the corporate world: IBM, like several other companies that offer a cloud-based IoT platform, operates a lab specifically dedicated to pen-testing of IoT systems; the company claims that its laboratory has discovered over 1000 new vulnerabilities since 2017 \cite{ibm2019}. While databases such as the National Vulnerability Database (NVD) keep track of IoT/CPS device vulnerabilities, the sheer speed with which new devices are being deployed has made up-to-date vulnerability cataloging all but impossible.

\par As a result, traditional exploit discovery and risk management engines are often incapable of properly modeling exploits in IoT/CPS networks. While some models have attempted to correct this problem by adapting the vulnerability scoring system or automatically-generating access conditions \cite{ur-rehman2019, ghazo2019}, none consider the unique (and often undiscovered) vulnerabilities of public-facing IoT/CPS devices or the convoluted exploits available to a clever adversary. Even ``premium" commercial risk management applications such as Tenable only examine risk from the perspective of individual vulnerabilities, rather than the chain of vulnerabilities through multiple devices that often constitute exploits in IoT/CPS \cite{tenable2021}. While some open-source risk management systems like MulVAL can theoretically find vulnerability chains in large systems, they do not model the undocumented (yet surprisingly common) vulnerabilities present in IoT/CPS and the topology-specific connections between them \cite{ou2005}.

\par GRAVITAS is also unique among IoT/CPS exploit discovery software in allowing the user to minimize the impact of a successful exploit and automatically optimize the placement of defenses to reduce risk at a minimum cost. Commercial software like Tenable claims to do risk optimization, but only does this at the vulnerability level and does not consider exploit chains between multiple devices \cite{tenable2021}. Other tools, such as TAG and VSA, claim to have optimization abilities, but disclose very little detail in their respective articles \cite{malowidzki2019, ur-rehman2019}. The information that does exist suggests that the available defenses for both these systems are few in number and extremely generic, bearing little resemblance to the transparent and mathematically-rigorous optimization algorithm applied by GRAVITAS. Table \ref{fig:comparison_chart} compares the features of GRAVITAS to similar exploit discovery and risk management software.

\begin {table}[h]
\centering
\caption {Exploit discovery model comparison} 
\label{fig:comparison_chart}
\scalebox{1}{
\begin{tabular}{|c|c|c|c|c|c|c|c|c|c|c|}
\hline
\textbf{Model} & \textbf{1} & \textbf{2} & \textbf{3} & \textbf{4} & \textbf{5} & \textbf{6} & \textbf{7} & \textbf{8} & \textbf{9} & \textbf{10} \\ \hline

GRAVITAS & \cmark & \cmark  & \cmark & \cmark & \cmark & \cmark & \cmark & \cmark & \cmark & \cmark \\ \hline
MulVAL \cite{ou2005} & \xmark & \cmark  & \xmark & \cmark & \xmark & \xmark & \cmark & \xmark & \xmark & \xmark \\ \hline
TVA \cite{jajodia2009}& \cmark & \xmark & \xmark & \cmark & \xmark & \xmark & \cmark & \xmark & \cmark & \xmark \\ \hline
TAG \cite{malowidzki2019} & \xmark & \cmark & \xmark & \cmark & \xmark & \cmark & \cmark & \cmark & \cmark & \xmark \\ \hline
A2G2V \cite{ghazo2019} & \xmark & \cmark & \xmark & \xmark & \xmark & \xmark & \cmark & \xmark & \xmark & \cmark \\ \hline
VSA \cite{ur-rehman2019} & \cmark & \xmark & \cmark & \xmark & \xmark & \cmark & \cmark & \xmark & \cmark & \cmark \\ \hline
Tenable \cite{tenable2021} & \xmark & \xmark & \cmark & \cmark & \cmark & \cmark & \cmark & \cmark & \cmark & \xmark \\ \hline
NetSPA \cite{lippmann2006} & \xmark & \cmark & \xmark & \cmark & \xmark & \xmark & \cmark & \xmark & \xmark & \xmark \\ \hline

\end{tabular}}
\end {table}
\small{The various columns in Table \ref{fig:comparison_chart} are described next.
\begin{enumerate}
    \item Includes vulnerabilities due to physical manipulation, as well as the hardware and software stacks.
    \item Discovers novel attack vectors.
    \item Calculates exploit risk based on experimentally-validated algorithm.
    \item Models different privileges within devices and their ability to access other devices.
    \item Incorporates novel attack vectors that have not yet been exploited in real-world systems.
    \item Optimizes defense placement to reduce exploit risk at minimum cost.
    \item Can find ``the weakest link" (most vulnerable part of the system).
    \item Accurately handles cyclic network topology.
    \item Allows easy customization based on a system administrator's risk impact assessment and chosen adversary model.
    \item Designed specifically for the unique characteristics of IoT/CPS networks.
    
\end{enumerate}}

\section{Background}
\label{section:Background}

GRAVITAS employs several concepts developed in previous exploit detection frameworks. These include the SHARKS IoT/CPS model, attack graphs, and the CVSS metrics for evaluating the impacts of vulnerabilities. This section 
provides an introduction to these concepts.

\subsection{The SHARKS Framework}

\begin{figure*}[t]
\centering
\includegraphics[width=0.85\linewidth,scale=0.95, height =0.52\linewidth]{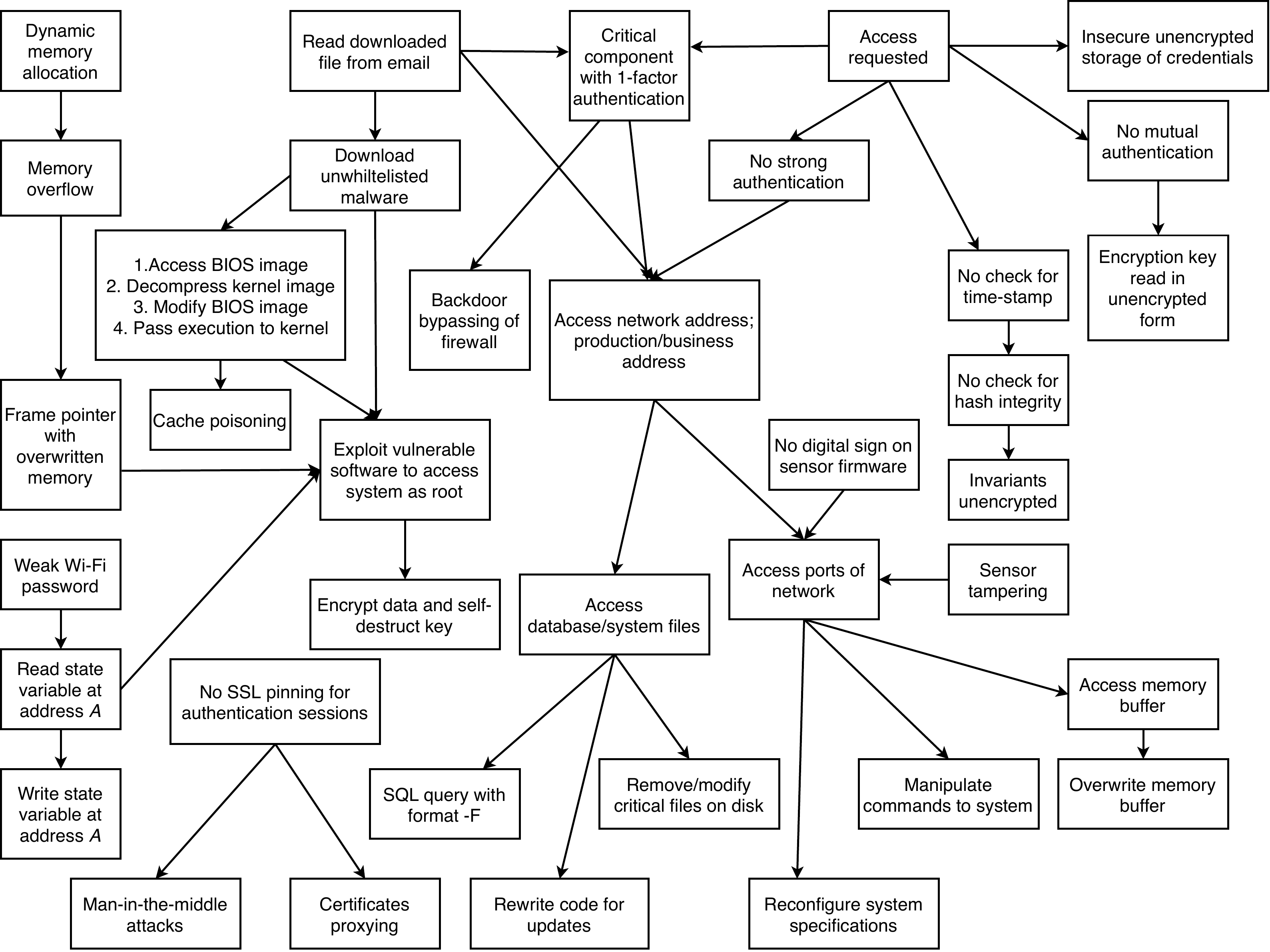}
\caption{The original SHARKS attack graph \cite{saha2020}} 
\label{fig:AttackDAG}
\end{figure*}

GRAVITAS builds on the work of SHARKS (Smart Hacking Approaches for RisK Scanning in
Internet-of-Things and Cyber-Physical Systems based on Machine Learning), which provides a novel framework for discovering IoT/CPS exploits \cite{saha2020}. Instead of artificially separating a system into different layers, SHARKS eschews a rigid classification and models an exploit chain (attack vector) as it appears to an adversary: a series of steps that begins at an ``entry point"  (a root node) and ends at a ``goal"  (a leaf node). The SHARKS attack graph (Fig.~\ref{fig:AttackDAG}) was created by deconstructing 41 known attacks on IoT/CPS into a series of steps represented by a vulnerability node chain (an exploit), and subsequently merging every node chain into a single directed acyclic graph (DAG). This graph makes no distinctions between network-level, hardware-level, or software-level nodes: what matters is the procedure that brings attackers to their desired destination.
 
\par SHARKS assigns descriptive features to each node in the attack graph, using one-hot encoding 
for categorical features as well as using continuous features such as the 
node's mean height in the graph. The authors assign a target feature to each pair of nodes 
indicating whether the two were connected by an edge (an ordered sequence of two steps in an attack). 
This dataset is then used to train a Support Vector Machine (SVM) model, which discovered 122 novel 
exploits. 

\subsection{Attack Graphs}

Both SHARKS and GRAVITAS are based on attack graphs, albeit with minor differences in their node types. We describe each graph ($G$) using the following terminology:

\begin{itemize}

\item \(N\): The set of nodes in the graph. Each node represents a single vulnerability in the system, such as ``sensor tampering" or ``no SSL pinning."
\item \(E\): The set of edges in the graph. Edges represent exploits, or pathways between vulnerabilities. Unlike in other attack graph models, edges do not have an 
access control parameter; each edge instead represents a possible path between vulnerabilities. 
Different permissions are instead represented by different nodes (see 
Section \ref{subsubsection:permission_subgraphs}).
\item \(D\): The set of nodes and edges corresponding to one device. Every device is depicted
by a subgraph of the complete attack graph (\(D \subset G\)).
\item $A$: The nodes at which an adversary can access the system. These ``entry nodes" are the starting points for any attack. They are also vulnerabilities ($A \subset N$).
\item $L$: The leaf nodes of the graph. These represent the completion of an attack ($L \subset N$). These ``exploit goals"  represent the end goal of an adversary's attack, such as "Disable device" or "DoS attack."
\item $P$: The set of nodes that constitute a complete attack vector. Each exploit begins at an entry 
node and concludes at an exploit goal node. More formally, an exploit $P$ is any ordered set of 
nodes in the form $\{ a, n_1, ..., n_k, l\}$ where $  a \in A $, $ n_i \in N $, $ l \in L $. The same entry node and exploit goal pair can be a part of multiple attack vectors.
\item $M$: The set of all defenses that can be applied to the attack graph.
\item $\widetilde{M}$: The subset of defenses chosen by the optimization process.
\end{itemize}

\subsection{Common Vulnerability Scoring System (CVSS)}

CVSS is an experimentally-validated scoring system for device vulnerabilities that is widely
used in risk management systems \cite{first2019specification}. In this article, every node in the attack graph is assigned a vulnerability score, which represents the intrinsic risk of the vulnerability to the system. This score incorporates both the impact of a vulnerability and its ease of exploitation, and is calculated using an algorithm almost identical to that used in CVSS to score vulnerabilities (see Section \ref{subsubsection:calculate_vulnerability_scores}). This algorithm uses the scoring factors described in Table \ref{fig:CVSS_scores}. There are three principal categories of 
factors: \textit{exploitability}, \textit{impact}, and \textit{defense}. \textit{Exploitability} refers to the effort required by an adversary to ``succeed" in an attack step, while \textit{impact} refers to the damage that a successful attack can inflict on the security of the system.
\textit{Defense} refers to the extent to which the attack is prevented from being exploited. The 
scores in the \textit{exploitability} category are determined by the composition of the attack graph 
and are computed algorithmically, while those for \textit{impact} are decided by 
the system administrator based on their judgement of an attack's impact. The \textit{defense} scores are hard-coded to the particular defense, though these can be modified by the user. In Table \ref{fig:CVSS_scores}, the scores marked with a * are taken from a study by Ur-Rehman et al. that adapts some of the original CVSS scores to better reflect the ease of exploitation of real-world IoT/CPS devices \cite{ur-rehman2019}. The scores marked with a $\dagger$ are newly-added scores that have been experimentally validated using the method described in Section \ref{subsection:parameter_validation}.

\begin {table}[h]
\caption {CVSS scoring values \cite{first2019specification}}
\begin{tabular}{|c|c|c|c|}
\hline
\textbf{Category} & \textbf{Factor} & \textbf{Type} & \textbf{Score}\\ \hline
\multirow{20}{*}{Exploitability} & \multirow{4}{*}{ \shortstack{Attack \\ vector}} & Network & 0.85 \\ \cline{3-4}
                                & & Adjacent & 0.62\\ \cline{3-4}
                                & & Local & $0.60^*$ \\ \cline{3-4}
                                & & Physical & $0.44^*$\\ \cline{2-4}
                                & \multirow{3}{*}{\shortstack{Attack \\ complexity}} & Low & 0.77 \\ \cline{3-4}
                                & & Medium & $0.44^*$  \\ \cline{3-4}
                                & & High & $0.20^*$ \\ \cline{2-4}
                                & \multirow{2}{*}{Scope} & Changed & N.A. \\ \cline{3-4}
                                & & Unchanged & N.A. \\ \cline{2-4}
                                & \multirow{5}{*}{\shortstack{Privileges \\ required}} & None & 0.85  \\ \cline{3-4}
                                & & Low; Scope Changed & 0.68 \\ \cline{3-4}
                                & & Low; Scope Unchanged & 0.62  \\ \cline{3-4}
                                & & High; Scope Changed & 0.50  \\ \cline{3-4}
                                & & Low; Scope Unchanged & 0.27  \\ \cline{2-4}
                                & \multirow{2}{*}{\shortstack{User \\ interaction}} & None & 0.85 \\ \cline{3-4}
                                & & Required & 0.62 \\ \cline{2-4}
                                & \multirow{4}{*}{\shortstack{Accessibility}} & High & $0.80^\dagger$ \\ \cline{3-4}
                                & & Medium & $0.60^\dagger$ \\ \cline{3-4}
                                & & Low & $0.40^\dagger$ \\ \cline{3-4}
                                & & None & $0^\dagger$ \\ \cline{1-4}
\multirow{3}{*}{Impact} & Confidentiality & High & 0.56 \\ \cline{3-4}
                                & Integrity & Low & 0.20\\ \cline{3-4}
                                & Availability & None & 0 \\ \cline{1-4}
\multirow{5}{*}{Defense}        & \multirow{3}{*}{\shortstack{\\ Node \& edge \\ defense}} & None & $1^\dagger$ \\ \cline{3-4}
                                & & Workaround & $0.90^\dagger$  \\ \cline{3-4}
                                & & Temporary & $0.60^\dagger$ \\ \cline{3-4}
                                & & Definite & $0.30^\dagger$ \\ \cline{3-4}
                                & & Infallible & $0^\dagger$ \\ \cline{1-4}
                                
\hline

\end{tabular}

\label{fig:CVSS_scores}		
\end {table}

\subsection{Threat Model}
\label{subsection:threat_model}

Our threat model consists of an adversary who wishes to achieve an exploit goal $l$ 
with a motivation specified by its \textit{impact} score. The adversary reaches 
the desired $l$ by starting at an entry node $a$ and passing through other vulnerability nodes 
$n_i \in N$. This complete path $P$ is known as an exploit chain or attack, and may involve 
vulnerabilities in multiple devices (see Section \ref{section:example_application} for an example). 
While the system administrator can assign a unique impact score to each $l$ 
or randomly assign the scores via a specified distribution, it is not known in 
advance which exact exploit goal an adversary may attempt to access. Hence, the administrator must 
take into account all possible exploits in the attack graph in order to minimize risk over the entire 
system. 	
		
\par For simplicity, we assume that the motivation of the adversary to inflict damage is congruent to the system administrator’s assessment of the damage that the attack would cause to system security. We also assume that the adversary also has access to GRAVITAS and may decide to perform an exploit based on information provided by the model. As a result, the system administrator has an incentive to ensure that high-impact exploits are not easily accessible to the adversary. This may require adding additional defenses to the system via the model's optimization component.

\section{Motivation}
\label{section:Motivation}
While SHARKS is able to find novel exploits on specific types of IoT/CPS, it is too generic to adequately model the multitude of intricate exploits in a real-world system. GRAVITAS solves this issue by creating a unique attack DAG for every device in an IoT/CPS and adding additional pathways between devices based on network topology. The goal is not to find specific vulnerabilities in each device (a process that often requires static and dynamic analysis of the system) but to understand the impact that an exploit would have on the security of the entire system if a certain vulnerability were to exist. Moreover, unlike other IoT/CPS security models, GRAVITAS can account for potential exploits that have not yet been discovered, enabling a proactive and comprehensive approach to network security that existing risk management tools cannot provide. GRAVITAS can also suggest defenses that reduce system vulnerability at the lowest cost, employing a ``defense-in-depth" optimization approach that is intractable to human analysis. By giving an administrator the ability to visualize and mitigate IoT/CPS exploits before the system is deployed, GRAVITAS hopes to prevent the next Mirai-like attack before it happens.

\section{Methodology}
\label{section::Methodology}

\begin{figure}[h!]
\centerline{\includegraphics[scale=0.48]{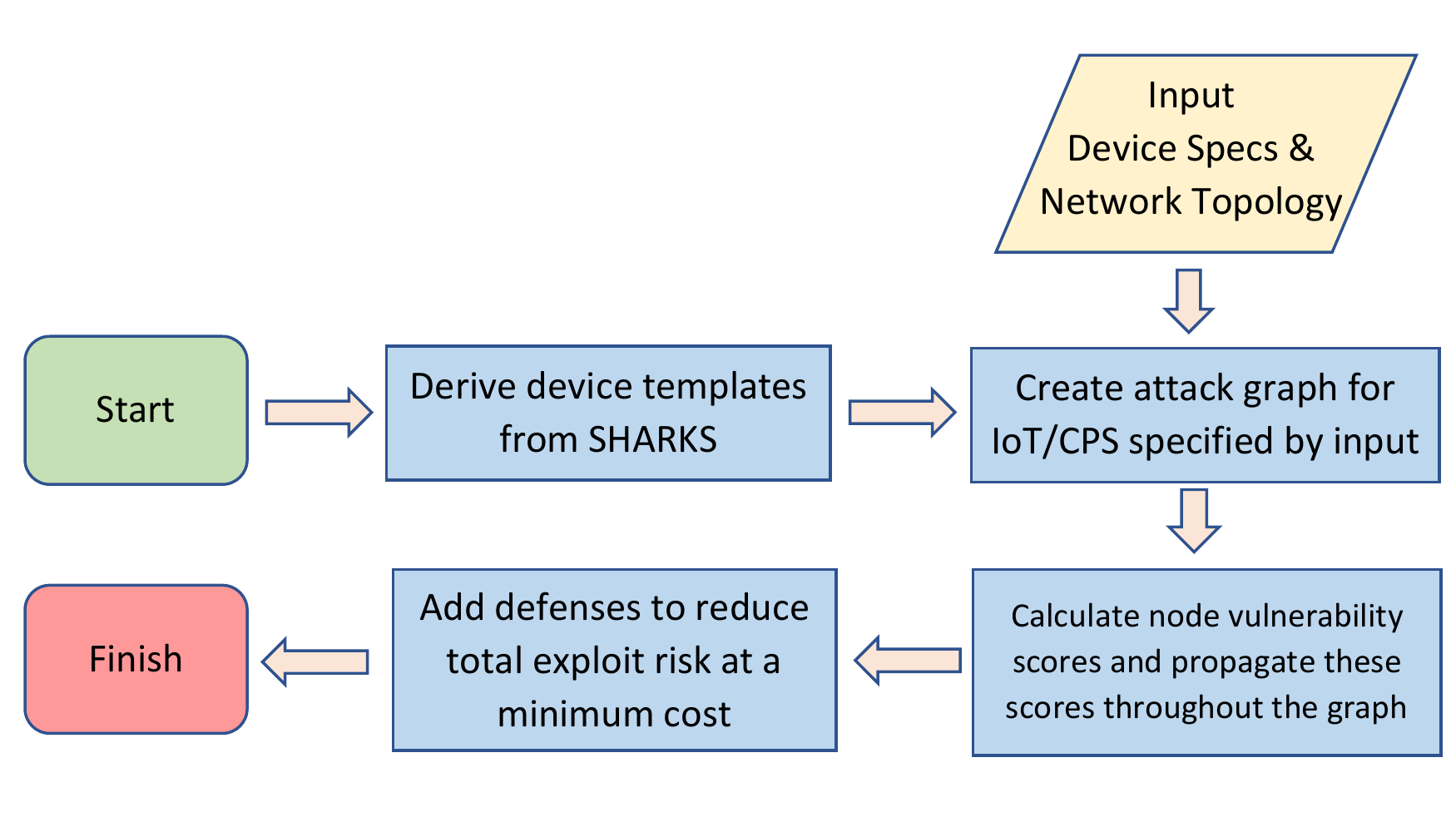}}
\caption{The overall structure of GRAVITAS}
\label{fig:overall_structure}
\end{figure}

In this section, we describe the GRAVITAS methodology.
As shown in Fig.~\ref{fig:overall_structure}, GRAVITAS consists of four primary components:
\begin{enumerate}
    \item Deriving the device templates from SHARKS.
    \item Creating an attack graph from the devices and network topology specified by the system administrator.
    \item Calculating the vulnerability score and exploit risk score for every node in the graph.
    \item Optimizing the placement of defenses to reduce the total exploit risk of the system.
\end{enumerate}

\begin{figure*}[t!]
\centering
\includegraphics[scale=0.85]{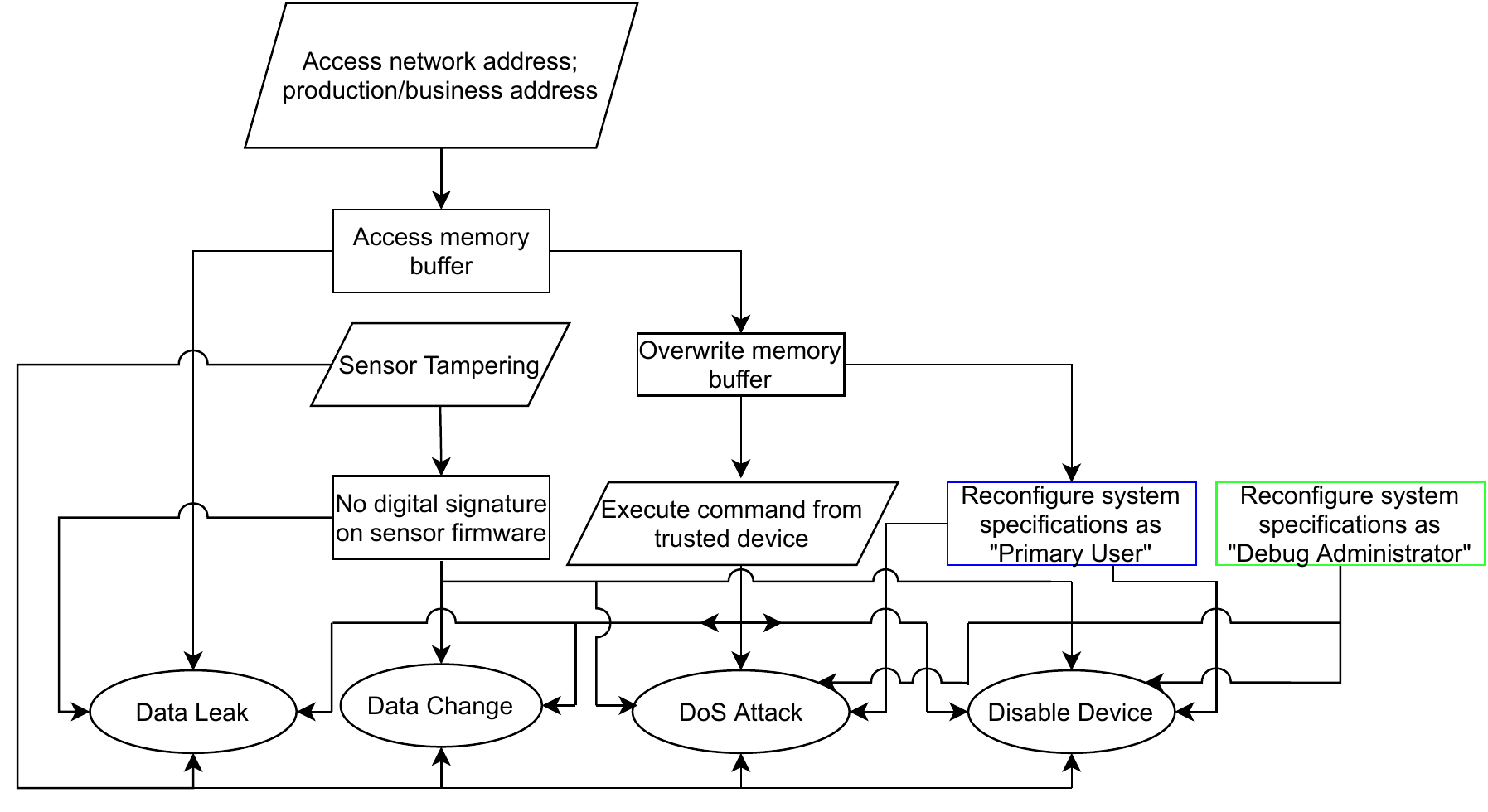}
\captionsetup{justification=raggedright, singlelinecheck=false}
\caption{Nest Garden Sensor in the smart home system as derived from a template for a Sensor (Non-updatable, Send and Receive, Local Network Access). Ovals represent exploit goals and parallelograms represent entry nodes. The blue and the green nodes represent the same subgraph that is repeated for each user-specified permission (see Section \ref{subsubsection:permission_subgraphs} for more detail).}
\label{fig:sensor_smarthome}
\end{figure*}

\begin {table*}[h!]
\caption {The exploit goals in the master attack graph template $J$}
\begin{tabular}{|m{4cm}|m{13cm}|}
\hline
\textbf{Exploit Goal} & \textbf{Example} \\ \hline
Eavesdropping over network & Unencrypted communication channel allows adversary to glean information \\ \hline
Denial-of-Service (DoS) & Mirai botnet used to flood DNS servers with bogus traffic \\ \hline
Disabling device & Adversary remotely turns off a smart city drone mid-operation, causing it to crash \\ \hline
Actuator malfunction & IoT-enabled pacemaker told to increase pace of electric shocks, causing serious harm to the patient \\ \hline
Data leak & Adversary accesses memory of a smart lock, learning the times of day when the victim is not at home \\ \hline
Data change & Adversary gains control of a traffic sensor and provides fake data to an ML algorithm that suggests routes for autonomous vehicles \\ \hline
Replay attack & Communications protocol does not use a digitally signed timestamp, allowing the adversary to resend previous commands \\ \hline
Ransomware attack & Adversary gains root access to device and encrypts essential files, asking the user to present payment in exchange for the key \\ \hline
Obtain authentication key to device $i$ with permissions $j$ & Adversary finds password on a
mobile device that permits him to login to a local controller from a different device \\ \hline
Obtain open access to device $i$ with permissions $j$ & Adversary finds password on a mobile
device that permits him to login to a local controller directly from the mobile device \\
\hline
\end{tabular}
\label{table:attack_outcomes}		
\end {table*}

These components are described in Sections \ref{subsection:derive_device_templates} through 
\ref{subsection:optimize_defense_placement}, respectively. Section 
\ref{subsubsection:adversary_models} describes the adversary model, whereas Section 
\ref{subsection:parameter_validation} describes a quality assurance program that employs 
randomly-generated smart city IoT/CPS to test the robustness of the internal parameters of
GRAVITAS.

\begin{figure*}[t!]
\centerline{\includegraphics[scale=0.75]{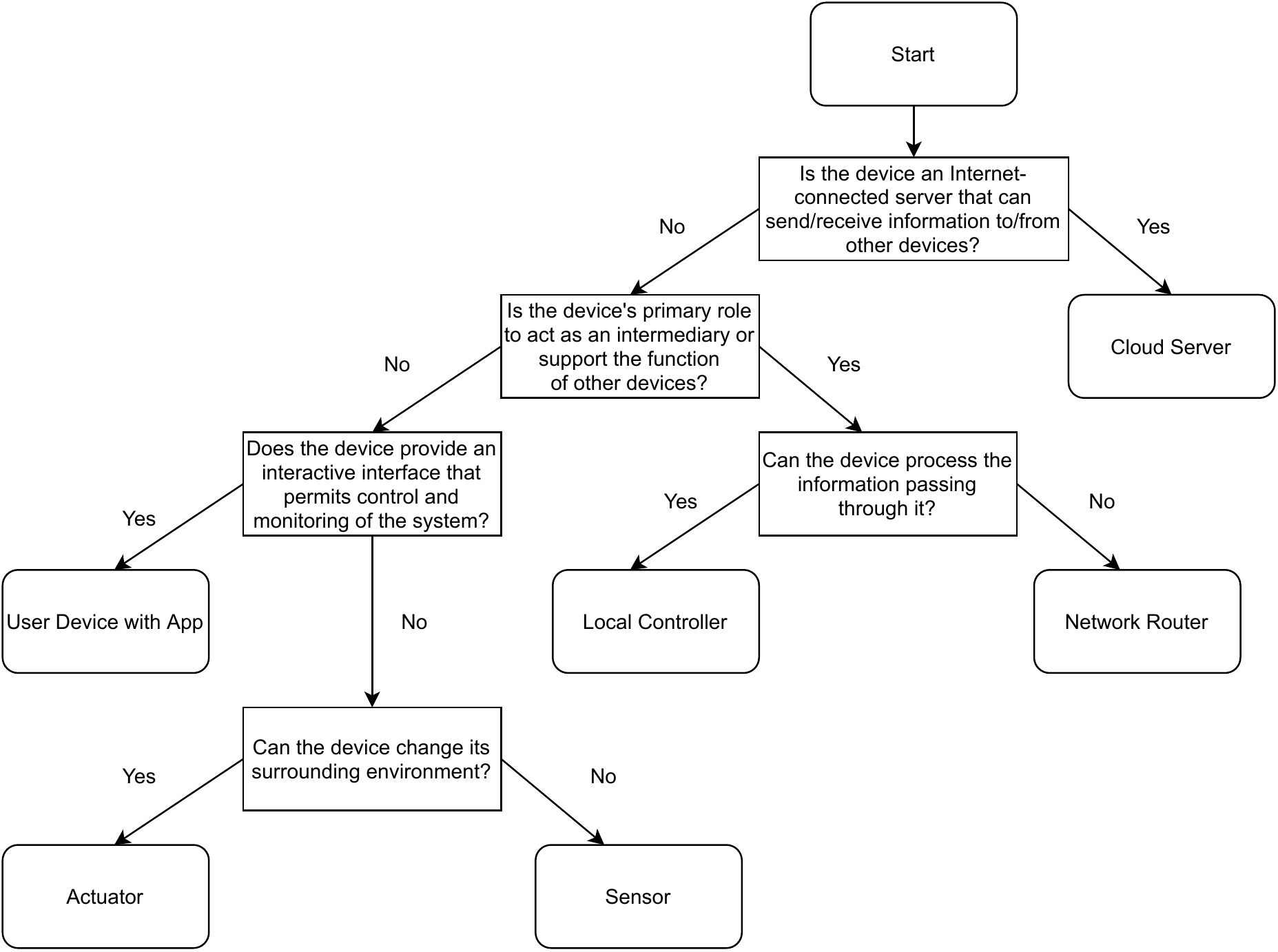}}
\caption{Flowchart for choosing the correct device category}
\label{fig:device_category_flowchart}
\vspace{12pt}

\small
\end{figure*}

\begin{table*}[h!]
\centering
\caption{Device subcategories for template construction}
\scalebox{0.9}{
\begin{tabular}{|c|m{2.1cm}|m{7.3cm}|m{7.3cm}|}

\hline
\textbf{Factor} & \textbf{Subcategory} & \textbf{Description} & \textbf{Examples} \\ \hline

\multirow{2}{*}{1} & Non-updatable & Device's application-level software and firmware cannot be changed & Devices with application software as immutable firmware, devices without a CPU (including some ASICs), devices without transistors or memory\\ \cline{2-4}
& Updatable & Device's application-level software or firmware can be changed & Any device that runs a version of Linux or most versions of RIoT, PSoC, and FPGA-like devices \\ \cline{3-4}
\hline
                                
\multirow{2}{*}{2} & Local network access & Device is only connected to a local network (no
direct Internet access) & Fitbit Versa 2 (still considered local network access only if
Internet-based data are accessed via a non-router proxy device) \\ \cline{2-4}
                    & External network access & Device is connected to the Internet (this can be through an adjacent router) & Apple Watch Series 5 Cellular Model \\ \cline{3-4}
                    \hline
                    
\multirow{3}{*}{3} & Send & Device is physically capable of broadcasting a signal in a format readable to other devices & A sensor with an output port containing a pin that can vary its voltage \\ \cline{2-4}
                                & Receive & Device is physically capable of receiving and 
``understanding" a signal from another device & An actuator with a simple electronic circuit
that depends on the input voltage level \\ \cline{2-4}
                                & Send and receive & Device can both send and receive information & 4G-enabled smartphone, drone with camera \\ \cline{1-4}
\hline

\end{tabular}}
\label{table:device_template_subcategories}
\end{table*}

\subsection{Deriving Device Templates}
\label{subsection:derive_device_templates}

\par All device attack graphs used in GRAVITAS are derived in part from an updated version of the SHARKS graph. This master attack graph template ($J$) consists of a subset of the original SHARKS graph, including ML-predicted edges between nodes that indicate new vulnerabilities. Each of these templates contains all known and predicted vulnerabilities and attack vectors for a given device. By including predicted exploits rather than just publicly known exploits in the system attack graph, GRAVITAS allows the system to be protected against possible novel exploits.
		
\par Every graph $J$ also contains a new set of nodes designated as exploit goals, $L$. Table \ref{table:attack_outcomes} lists the exploit goals; they collectively represent all of the IoT/CPS exploits described in \cite{mosenia2017comprehensive}. In addition, we designate certain nodes from the original SHARKS graph as entry nodes, $A$. Even with these additions, the master attack graph $J$ is still a DAG, ensuring that any exploit $P$ derived from it will be finite in length and non-repeating. GRAVITAS contains dozens of different templates for IoT/CPS devices, each of which is further customizable based on user input. Fig. \ref{fig:sensor_smarthome} shows an example of a device template that has been customized to a device present in the smart home example application (see Section \ref{section:example_application} for more detail).

\subsubsection{Creating the Device Templates}
\label{subsubsection:create_the_device_templates}

Every device template $T_i$ is a subgraph of $J$, the master attack graph ($T_i \subset J$). A device 
graph $D$ consists of the corresponding device template with modifications specified by the system administrator's 
input. While each device graph $D$ derived from a template is a DAG, the complete system attack graph 
$G$ may contain cycles in certain network topologies. This is acceptable because the ``long" cycles 
created by pathways between devices still permit the exploit risk score calculation 
process to converge in a reasonable time frame (see Section \ref{subsubsection:propagate_exploit_risk_scores} for 
further details). 

\par To create a device graph $D$, we must first classify the device
using four factors. The first of these factors, which we refer to as a \textit{category}, 
describes the purpose of the device; Fig.~\ref{fig:device_category_flowchart} shows a flowchart
that helps the system administrator decide which of the six categories to select. We refer to the other three factors
as \textit{subcategories}, each of which describes a physical limitation of the device and
consequently a hard boundary on the kind of attacks it is susceptible to. These subcategories
are described in Table \ref{table:device_template_subcategories}. The categories and
subcategories together provide comprehensive coverage of the numerous IoT/CPS applications
listed in \cite{mosenia2017comprehensive}.

\subsection{Creating IoT/CPS Designs}
\label{section:create_userspecified_IoT_system}

\begin{figure}[h]
\centering
\includegraphics[scale=0.7]{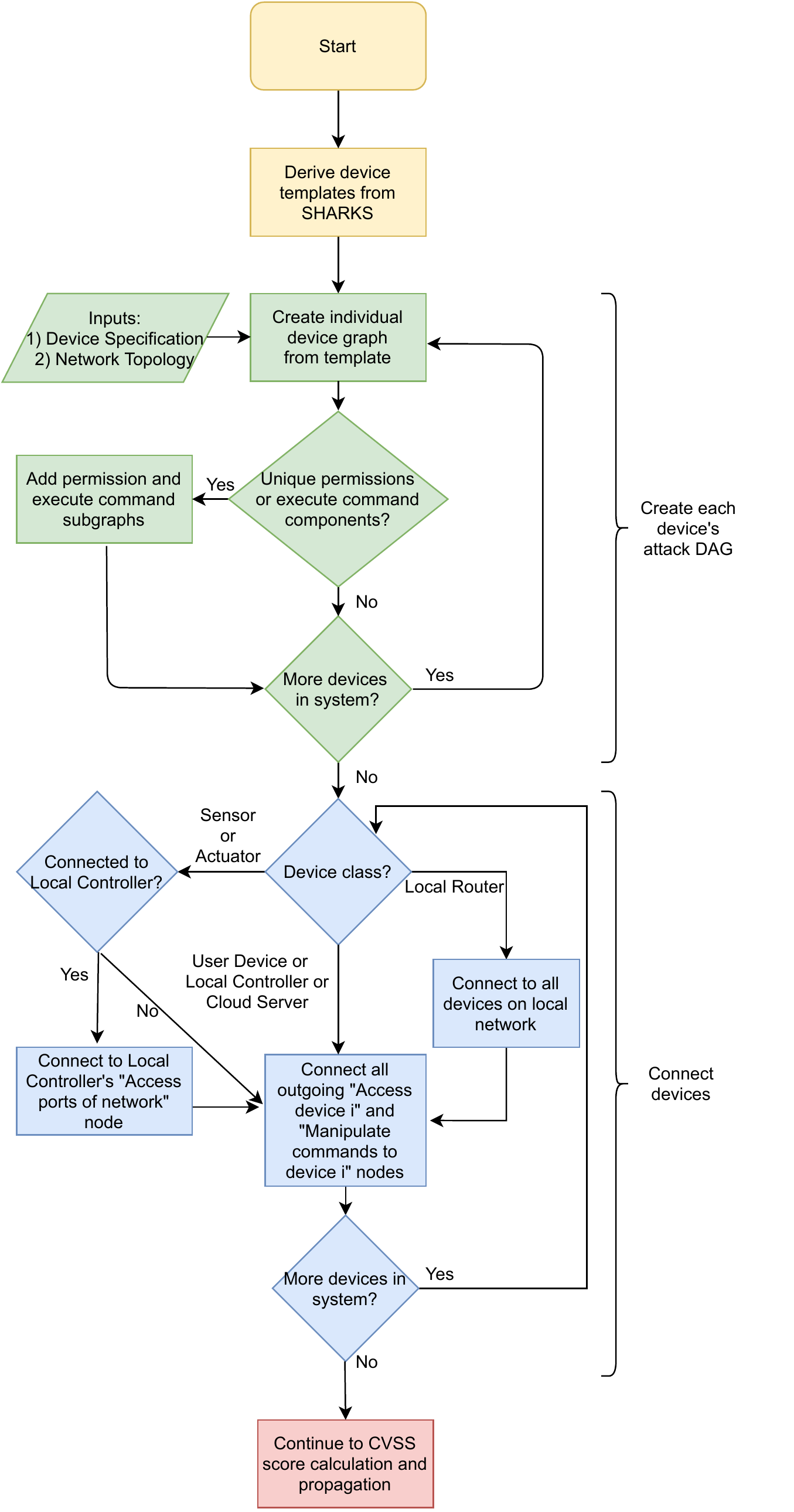}
\caption{The procedure for creating the complete system graph $G$ from system administrator input} 
\label{fig:create_IoT_system}
\end{figure}

Fig.~\ref{fig:create_IoT_system} provides an overview of how GRAVITAS creates an attack graph $G$ 
from system administrator input. In addition to specifying inherent device properties, such as the device category 
and subcategories, the system administrator must also describe each device's location in the network topology. This
includes its connections (wired and wireless) to devices in the local network as well as its ability 
to connect to an external network such as the Internet. Table
\ref{fig:userspecified_device_properties} describes a subset of these device-level
characteristics. Note that most of these characteristics are entirely optional. Only ``Name," ``Category," and ``Subcategory" are strictly necessary; all other characteristics receive default values that treat the device as a 
disconnected component with a low security risk. If the system administrator is unsure about the topology of a certain device or its \textit{impact} scores, GRAVITAS can automatically generate the connections and \textit{impact} scores from a random distribution with parameters set by the administrator. For an example of a system design, see 
Section \ref{section:example_application}.

\begin {table}[h]
\caption {Administrator-specified device properties}
\begin{tabular}{|m{2cm}|m{6cm}|}
\hline
\textbf{Device property} & \textbf{Description} \\ \hline
Name & Unique name of device \\ \hline
Category & Choose from the categories in Fig.~\ref{fig:device_category_flowchart}	\\ \hline
Subcategory & Choose from the subcategories in Table \ref{table:device_template_subcategories} \\ \hline
Device set & Devices in the same set are treated as identical; during system optimization, all devices in a single set are given the same defense concurrently\\ \hline
Confidentiality scores & A list of CVSS confidentiality scores for the exploit goals of this device\\ \hline
Integrity scores & A list of CVSS integrity scores for the exploit goals of this device\\ \hline
Availability scores & A list of CVSS availability scores for the exploit goals of this device\\ \hline
Accessibility scores & A list of accessibility scores for the entry nodes of this device\\ \hline
Login permissions & A list of devices that possess credentials to login to this
device; a single device may have multiple permission types, and multiple devices may login to this device using the same permission type\\ \hline 
Execute command permissions & A list of devices that possess credentials to send
valid commands to this device; a single device may have multiple permission types, and multiple devices may login to this device using the same permission type\\ \hline
General permissions & Some simple devices, such as embedded IoT sensors, do not possess a formal security protocol, and can be controlled by any device that connects to them \\ \hline
Local networked devices & A list of devices in this device's local network; if this is a
``hub"  network centered around a device like a WiFi router, the algorithm will automatically connect 
all devices in a local network group as long as each device is connected to the hub \\ \hline
\end{tabular}
\label{fig:userspecified_device_properties}		
\end {table}

\subsubsection{Defenses}
\label{subsubsection:user_input}

\begin {table}[h]
\caption {Defense properties}
\begin{tabular}{|m{1.5cm}|m{6.5cm}|}
\hline
\textbf{Defense property} & \textbf{Description} \\ \hline
Defense name & Unique name for the defense; all defenses in a device set that share the same name will be applied simultaneously if chosen during optimization \\ \hline
Device name & A list of devices to which this defense can be applied	\\ \hline
Cost & The cost of defense in relative units \\ \hline
Node score list & A list containing each node and updated \textit{node defense score} affected by the defense \\ \hline
Edge score list & A list containing each edge and updated \textit{edge defense score} affected by the defense \\ \hline
\end{tabular}
\label{fig:defense_properties}		
\end {table}

\begin {table}[h]
\caption {Selected defenses in $M$ (with default settings)}
\begin{tabular}{|m{3cm}|m{0.8cm}|m{2.9cm}|m{0.6cm}|}
\hline
\textbf{Defense} & \textbf{Type} & \textbf{Vulnerability(-ies)} & \textbf{Cost} \\ \hline
Multifactor authentication & Single node & No mutual authentication & 2 \\ \hline
Salt, hash, and encrypt passwords & Single node & Encryption key read from memory in unencrypted form & 1 \\ \hline
Limit installation of unverified software & Single node & Download unwhitelisted malware & 1 \\ \hline
Limit physical access to device ports & Multi node & Sensor tampering, side-channel analysis & 3 \\ \hline
Establish difficult-to-crack device passwords & Edge & Access requested, no strong authentication & 2 \\ \hline

\end{tabular}
\label{fig:selected_defenses}		
\end {table}

GRAVITAS provides $M$, the set of potential defenses that can be applied during optimization. The defenses in GRAVITAS are derived from ENISA's \textit{Baseline Security Recommendations for IoT}  \cite{enisa2017}. This list includes vulnerability mitigation strategies identified in NIST, OWASP, ISO, and a variety of other standard-setting organizations. Similar to the widely-employed mitigation techniques cataloged in MITRE’s ATT\&CK database \cite{mitre2021}, the defenses employed by GRAVITAS correspond to a vulnerability category (“technique") that is present across devices in the same category. Section \ref{section:example_application} illustrates how an optimized subset of these defenses can be applied to exploits in a real-world system. Note that the system administrator can add or subtract defenses to $M$ if additional specificity or generality is required, and also change the defense cost.

Table \ref{fig:defense_properties} describes the properties of a defense in detail. Defenses can act at two places: nodes and edges. A 
\textit{node defense} changes the score used in the node’s vulnerability score calculation 
(see Section \ref{subsubsection:calculate_vulnerability_scores}). An \textit{edge defense} 
changes the weight multiplier of the corresponding edge during exploit risk score 
propagation (see Section \ref{subsubsection:propagate_exploit_risk_scores}). The cost of 
each defense and its marginal improvement to the total exploit risk (adversary score) are used in the objective function 
that determines which defense to add (see Section \ref{subsubsection:objective_functions}). Table \ref{fig:selected_defenses} shows a small selection of node (vulnerability) defenses in $M$.

\par Every device graph $D$ initially gives each of the device's permissions blanket access to a device's capabilities, similar to administrator privileges. To limit the access of a certain permission, the system administrator can specify a defense that removes an edge between nodes in a permission subgraph (see Section \ref{subsubsection:permission_subgraphs}). This blacklist approach simplifies the input and ensures that low-cost/high-impact defenses are added at the beginning of the optimization process.	
		
In GRAVITAS, the cost of each defense is fixed at the outset of optimization. However, system administrators can set their own cost values based on various constraints of the system such as latency, expenditure, and energy. Because the cost and adversary score are weighted using a 
constant in the optimization function, the administrator can think of the cost of each defense in 
relative terms when deciding which values to choose. As a result, the specific numbers are not as 
important as the relative cost between different kinds of defenses: for example, the administrator 
could classify a defense’s cost as either ``high," ``medium," or ``low," each of which has a specific 
numerical value associated with it in a manner similar to a CVSS scoring attribute.

\subsubsection{Permission Subgraphs}
\label{subsubsection:permission_subgraphs}

GRAVITAS allows the system administrator to specify permissions for every device. Unlike other attack graph models, the access permissions are each represented by a separate copy of a subgraph rather than as a logical statement at certain nodes or edges \cite{kordy2014, ou2012, noel2010}. This design choice makes a visualization of the attack graph easier to understand, and also simplifies the calculation of exploit risk scores because we can apply the same closed-form calculation to every node-edge pair (see Section \ref{subsubsection:propagate_exploit_risk_scores}).

\par We model two different types of permissions: login permissions and execute command permissions. 
With a login permission, a user with the correct credentials can execute any (permitted) command on 
the system; this is similar to a user profile on a Linux or Windows system. With an execute command 
permission, a user with the correct credentials can execute a (permitted) command from a specified 
list. For example, this could be a set of commands recognizable in a JavaScript Object Notation (JSON) packet or the movement controls for an autonomous drone. Depending on the configuration of the IoT/CPS, different devices may have login/execute commands under the same permission name. 

\par As described in Table \ref{fig:permission_subgraph_node_types}, certain nodes in every device 
template are associated with login permissions, execute command permissions, or both. These subgraphs 
are ``repeated"  for every permission type of that device.   Fig.~\ref{fig:sensor_smarthome} shows the system-specific device graph 
$D$ with nodes in different permission subgraphs delineated in different colors. Note that the system administrator can specify defenses that effectively remove certain nodes or edges, allowing the administrator to set restrictions on what each permission can access.

\begin {table}[h]
\caption {Permission subgraph node types}
\begin{tabular}{|m{4.7cm}|m{1.5cm}|m{1.5cm}|}
\hline
\textbf{Attack node} & \textbf{Login permission} & \textbf{Execute command permission} \\ \hline
Access network address; production/business address & Yes & No \\ \hline
Access ports of network & Yes & No \\ \hline
Reconfigure system specifications & Yes & Yes \\ \hline
Access database/system files & Yes & Yes \\ \hline
SQL query with format -F & Yes & Yes \\ \hline
Rewrite code for updates & Yes & Yes \\ \hline
Remove/modify files on disk & Yes & Yes \\ \hline
Download unwhitelisted malware & Yes & Yes \\ \hline
Read state variable at address A & Yes & Yes \\ \hline
Write state variable at address A & Yes & Yes \\ \hline
Execute command & No & Yes \\ \hline
\end{tabular}
\label{fig:permission_subgraph_node_types}		
\end {table}

\subsubsection{Connecting the Devices}
\label{subsubsection:connect_the_devices}

Once every device's attack graph $D$ has been produced, we can connect them together into an
aggregate attack graph $G$ using the network topology specified by the system administrator.
A login permission 
$j$ is represented by an edge originating from the exploit goal ``Obtain authentication key to 
device $i$ as permission $j$," while an execute command permission $j$ is represented by an edge 
originating from ``Manipulate commands to device $i$ with permission $j$." Both lead to the 
node ``Access network address, production/business address as permission $j$" of device $i$. Both of 
these connections allow the devices to bypass the device's authentication procedures, meaning that 
the attacker does not have to start at the ``Access requested" node where almost all non-authenticated 
adversaries must begin their attack. For certain sensors and actuators, ``Sensor tampering" and 
``No digital signature on sensor firmware" are connected to the ``Access ports of network" node of the 
neighboring local controller; this represents the local network that exists between certain local 
controller and sensor/actuator setups, such as those involving an Arduino. To model access to a local 
network, every router's ``No strong authentication" node is connected in both directions to that same 
node in all other adjacent routers, and is also connected to the ``Access requested"  node for 
adjacent devices that are not routers. External network access (i.e., to the Internet) is modeled by 
including nodes such as ``Download unwhitelisted malware" in the device template.

\subsection{Calculating and Propagating Vulnerability Scores}
\label{subsection:calculate_and_propagate_vulnerability_scores}

Every node in the attack graph is first assigned a vulnerability score. These scores are then ``propagated" through the graph, giving each attack node an exploit risk score. The total exploit risk of the IoT system or CPS, which we call the \textit{adversary score}, is calculated using the adversary model chosen by the system administrator and involves a function of the exploit risk scores of entry nodes (see Section \ref{subsubsection:adversary_models}). This score is used in the objective function employed in the defense placement optimization process described in Section \ref{subsection:optimize_defense_placement}. The vulnerability scores are only calculated once, whereas the exploit risk scores (and adversary score) must be recalculated after adding a new defense.

\par All vulnerability and exploit risk scores fall into the [0,1] range. This is a departure 
from the traditional CVSS scoring range of [0,10], but it allows us to treat each score as the 
probability that an adversary will attempt the attack and succeed in exploiting it. This approach is 
widely used in attack graph models because it allows for a probabilistic understanding of an 
adversary's movement through the graph \cite{noel2010, aksu2017, ur-rehman2019}.

\subsubsection{Calculating vulnerability scores}
\label{subsubsection:calculate_vulnerability_scores}
\setlength{\algomargin}{0pt} 
\begin{algorithm}[h]
\small
\SetAlgoLined

$ISS = 1-[(1-confidentiality)(1-integrity)(1-availability)(1-accessibility)]$\;

\eIf{$scope$ \emph{is} unchanged}
{$Impact = 6.42 \times ISS$\;}
{ $Impact = 7.52 \times (ISS-0.029) - 3.25 \times (ISS-0.02)^{15} $\;}

$Exploitability = 8.22 \times attack\_Vector \times attack\_Complexity \times privileges\_Required \times user\_Interaction$\;
\
\If{$Impact < 0$}
{$Impact = 0$\;}

\eIf{$scope$ \emph{is} unchanged}{$x = min[(Impact + Exploitability), 10]$\;}
{$x = min[1.08 \times (Impact + Exploitability), 10]$\;}

$vulnerability\_Score = \frac{x}{10} \times defense\_Score$

\caption{Vulnerability score calculation}
\label{fig:vulnerability_score_calculation}
\end{algorithm}

Algorithm \ref{fig:vulnerability_score_calculation} describes the process for calculating the
vulnerability score. This algorithm (including its constants) is identical to the 
experimentally-validated ``Base Metrics Equations" present on page 18 of the CVSS Specification 
Guide, with the addition of a final line that adjusts the vulnerability score based on the 
defense(s) applied to the system \cite{first2019specification}.

Each node's vulnerability score is calculated using the \textit{Exploitability} factors ($scope$, $attack\_Vector$, $attack\_Complexity$, and $user\_Interaction$) described in 
Table \ref{fig:CVSS_scores}. The value of each factor is computationally 
determined by an algorithmic version of the corresponding flowchart on pages 20-21 of the CVSS 
v3.1 user guide, and is thus dependent on the node's permission characteristics and its topographic 
position within the device graph $D$ \cite{first2019userguide}. The \textit{impact} factors ($confidentiality$, $integrity$, and 
$availability$) are set by the system administrator, as is the $accessibility$ score of 
each entry node. These factors are used to calculate the $ISS$ (Impact Sub-Score). If the system administrator does not want to specify the impact score for every exploit goal, the scores can be auto-generated by providing parameters for the distribution of impact scores in the specified system components. This process is similar to the one used to randomly generate IoT/CPS for testing purposes (see Section \ref{subsection:parameter_validation} for more detail).

\subsubsection{Propagating Exploit Risk Scores}
\label{subsubsection:propagate_exploit_risk_scores}

To model IoT/CPS security, we need an understanding of how vulnerability scores of 
different nodes interact. This interaction is represented by each node's exploit risk
score, which is calculated using a function that involves the exploit risk scores of 
adjacent child nodes (see Algorithm \ref{fig:node_algorithm}). 

\par An adversary generally wants to take the least risky path possible through the attack graph; 
consequently, the longer and more difficult the path, the less likely the adversary is to pursue it,
and the lower the exploit risk score should be. In a similar fashion, we would consider an 
attack graph to be more vulnerable if there are multiple paths to the same exploit goal. To represent this 
idea in the exploit risk score calculation (Algorithm \ref{fig:node_algorithm}), we employ a probabilistic union function that 
gives the likelihood that an adversary will execute at least one of the exploits flowing from the current node to its children \cite{aksu2017}. This method is effective because it incorporates the
children nodes' exploit risk scores (and the scores of their descendants, including 
exploit goals) into the exploit risk scores of their parents. Consequently, the exploit risk score of an entry node represents the vulnerability of all exploits
reachable from that location. Section \ref{subsection:parameter_validation} provides a description of how the parameters and functions in this algorithm were experimentally determined.

\begin{algorithm}[h]
\small
\SetKwFunction{FMain}{$calculate\_Exploit\_Score$}
  \SetKwProg{Fn}{Function}{:}{}
  \Fn{\FMain{$n$, $G$}}{
\KwData{Node $n$; Attack graph $G$}
\KwResult{Exploit risk score for $n$}
\CommentSty{// $c_i$ is a child of node $n$}\label{cmt}
$union\_Probability = n.vulnerability\_Score \times (1 - \prod_{i=1}^{k}(1-(c_i\times edge\_Defense(n, c_i))))$\;
$exploit\_Score = -7^{-union\_Probability}+1$ \;
\CommentSty{/* Activation function (see Section \ref{subsection:parameter_validation} for more details about how this function was determined) */}\label{cmt}

\If{$exploit\_Score > 1$}{$exploit\_Score = 1$\;}
\If{$exploit\_Score < 0$}{$exploit\_Score = 0$\;}

\KwRet{exploit\_Score}\;}
\caption{Exploit risk calculation}
\label{fig:node_algorithm}
\end{algorithm}

\begin{algorithm}[h]
\small
\KwData{Attack graph $G$ with nodes $n_1$, ... ,$n_k \in N$; entry nodes $A \displaystyle \subset N$; exploit goal nodes $L \displaystyle \subset N$} 
\KwResult{Attack graph $G$ with exploit risk scores at each $n_i$}
$average\_Sum = $ $\mid A \mid$\;
\While{$\frac{average\_Sum}{\mid A \mid} > sum\_Ratio$ \emph{and}  $count < max\_Cycle$}{
    $average\_Sum = 0$\;
    Queue $Q \longleftarrow L$\; 
    \While{$\mid Q \mid$ $ > 0$}{
        $n \leftarrow Q$.dequeue\;
        \If{$n$ \emph{has not been visited} }{
            $n.exploit\_Score =$ \textbf{calculate\_Exploit\_Score(}$n$, $G$) \;
            
            \If{n $\in$ A}{
                   $average\_Sum = average\_Sum + | n.exploit\_Score - n.exp\_Avg | $\;
            }
            
            $n.exp\_Avg = exp\_Weight \times n.exploit\_Score + (1-exp\_Weight) \times n.exp\_Avg$\;
            Mark $n$ as visited\;
            
            \For{$parent\_Node$ \emph{in} $n.parents$}{
                \If{$parent\_Node$ \emph{has not been visited}}{
                    $Q$.append($parent\_Node$)\;
                }
            }
        }
    }
    $count+=1$\;
}
\caption{Propagating exploit risk scores}
\label{fig:exploit_risk_score_algorithm}
\end{algorithm}

\par If we think of each vulnerability as a neuron, we can think of the attack graph as a recurrent neural network (RNN), where the neuron-level score 
calculation combines inputs from multiple adjacent children nodes that are fed into an activation 
function. As with forward propagation in an RNN, our score propagation algorithm uses a 
breadth-first search that halts at nodes that have already been visited; once all nodes have been 
visited, we repeat the whole process and continue this repetition until the scores at the entry nodes 
have converged. Algorithm \ref{fig:exploit_risk_score_algorithm} shows this process in detail.

\begin {table*}[h]
\caption {Administrator-specified parameters for controlling the score propagation and optimization process}
\begin{tabular}{|l|m{4cm}|m{8cm}|m{2.5cm}|}
\hline
\textbf{Parameter} & \textbf{Purpose} & \textbf{Description} & \textbf{Suggested values} \\ \hline
$sum\_Ratio$ & Limits number of propagation cycles in Algorithm \ref{fig:exploit_risk_score_algorithm} to reduce computation time & Sets a floor on the average of the difference between every entry node's most recent exploit score and a weighted average of its exploit scores from previous propagation cycles. This difference should approach zero as the exploit scores converge. & $10^{-4}$ to $10^{-6}$  \\ \hline
$exp\_Weight$ & Helps measure progress towards the convergence of entry node exploit scores & This is the parameter for determining the exponentially-weighted moving average of a node's exploit risk score over all propagation cycles. We compare this average to the newest score to gauge convergence. & $0.05$ to $0.2$ \\ \hline
$max\_Cycle$ & Limits number of propagation cycles in Algorithm \ref{fig:exploit_risk_score_algorithm} to reduce computation time & Sets a ceiling on the number of propagation cycles & $50$ to $250$ \\ \hline
$max\_Defenses$ & Limits number of optimization rounds & The maximum number of defenses that can be added during optimization &  $|N| \times $[0.05 to 0.2] \\ \hline
$opt\_Halt\_Value$ & Limits the number of optimization rounds & Sets a floor for the global
objective value, below which optimization ceases & Preference-dependent \\ \hline
$\alpha\_Global$ & The weighting for the global objective function & The weighting of
$total\_Cost$ against $adversary\_Score$ increases linearly with $\alpha\_Global$ & Preference-dependent \\ \hline 
$\alpha\_Local$ & The weighting for the local objective function & The weighting of $defense\_Cost$ against the marginal difference in $adversary\_Score$ increases linearly with $\alpha\_Local$ & $0$ to $0.0001$ \\ \hline
$max\_Set\_Time$ & Improve optimization performance by removing low-performing defenses  & The maximum amount of rounds that a defense may remain in the defense set & $10$ to $50$  \\ \hline
$set\_Size$ & Reduce computation time by comparing less defenses in every optimization round & The number of defenses in the defense set. A larger defense set will result in a better optimization result & $0.05 \times |D|$ at minimum \\ \hline

\end{tabular}
\label{table:algorithm_parameters}		
\end {table*}

\subsection{Optimizing Defense Placement}
\label{subsection:optimize_defense_placement}

\begin{algorithm}
\small
\KwData{Attack graph $G$; defenses $d_1$, ... ,$d_j \subset M$}

\KwResult{$H$, $h\_Best$; \CommentSty{/* $h\_Best$ includes the optimally-defended graph $G$ and optimal defense set $\widetilde{M} \subset M$} */}

 $S$ = \textbf{refresh\_Defense\_Set}($\emptyset$, $G$)\CommentSty{; /* $S$ is a set containing graphs that each differ in only one defense */}

$H$ $=$ $\emptyset$\;

\Repeat{$\mid S \mid$ $\leq 0$ \emph{and} $\mid H \mid$ $\geq max\_Defenses$ \emph{and} \emph{\textbf{global\_Objective\_Func}}($\alpha\_Global$\emph{, total\_Cost(}$h.\widetilde{M}$\emph{), adversary\_Score(}$h.G$\emph{)) }$\leq opt\_Halt\_Value$ }{

    $chosen\_Min\_Obj\_Val = \infty$\;
    \For{$s$ \emph{in} $S$}{
    
        $local\_Obj\_Value = $ \textbf{local\_Objective\_Func}($\alpha\_Local$, $s.d.cost$,
adversary\_Score($s.G$))\;
        \If{$local\_Obj\_Value$ $< chosen\_Min\_Obj\_Val$}{
        
            $chosen\_Set = s$\;
        }
        $s.time\_In\_Set = s.time\_In\_Set + 1$\;
    }
    $total\_Cost = total\_Cost +chosen\_Set.d.cost$\;
    
    $H$.add(make\_History\_Moment($chosen\_Set.d$, $chosen\_Set.G$))\;
    
    $S$.remove($chosen\_Set$)\;
    
    $S = \textbf{refresh\_Defense\_Set}$($chosen\_Set.d$, $chosen\_Set.G$)\;
}

$min\_Global\_Obj\_Val = \infty$\;
\For{$h \in$ $H$}{
    
    $global\_Obj\_Val =$  \textbf{global\_Objective\_Func}($\alpha\_Global$, total\_Cost($h.\widetilde{M}$), adversary\_Score($h.G$))\;

        \If{$global\_Obj\_Val$ $< min\_Global\_Obj\_Val$}{
        
            $min\_Global\_Obj\_Val = global\_Obj\_Val$\;
            $h\_Best = h$
        }
}

\caption{Optimizing the system}
\label{fig:optimize_system}
    
\end{algorithm}

The optimization component of GRAVITAS adds defenses to the system with the goal of minimizing 
objective functions specified by the system administrator. It does this by creating a
``history"  set $H$ that records the defense selections of successive optimization rounds. One "moment" in history $h \in H$ refers to a single iteration of executing the local objective function; each $h$ includes the defense that was just chosen as well as an attack graph whose scores have been updated to include the just-chosen defense and all previously-added defenses. Once all $h$ have been determined, we use a global objective function to choose the optimal defense set $\widetilde{M}$ from among all $h$.
Algorithm \ref{fig:optimize_system} provides an overview of the optimization algorithm, whereas 
Algorithm \ref{fig:refresh_defense_set} describes the \textbf{refresh\_Defense\_Set}($d$, $G$) 
function that is used to generate the defense set $S$ whose hypothetical defense-graph pairings are compared by the local objective function. 
Section \ref{subsubsection:objective_functions} describes the local and global objective functions in 
detail. Table \ref{table:algorithm_parameters} describes the parameters the administrator can set to further refine the optimization process.

\subsubsection{Objective Functions}
\label{subsubsection:objective_functions}

\begin{flushleft} \small
Choose from all defense set members $s \in S$, where $s$ consists of a just-added defense $d$ and 
corresponding graph $G$.
\end{flushleft}
\begin{equation}
\label{eqn:local_function}
\begin{gathered}
\textbf{min}[\alpha\_Local \times s.d.cost+(1-\alpha\_Local) \\ \times (\emph{adversary\_Score}(h_{t-1}.G) - \emph{adversary\_Score}(s.G))]
\end{gathered}
\end{equation}

\begin{flushleft} \small
Choose from among all moments in history $h \in H$, where $h$ consists of defense set 
$\widetilde{M}$ and corresponding graph $G$ with all $d \in \widetilde{M}$ added.
\end{flushleft}
\begin{equation}
\label{eqn:global_function}
\begin{gathered}
\textbf{min}[\alpha\_Global \times total\_Cost( h.\widetilde{M}) \\ + (1-\alpha\_Global)\times \emph{adversary\_Score}(h.G)]
\end{gathered}
\end{equation}

The purpose of each objective function is to minimize the system's total exploit risk (\emph{adversary\_Score}) while simultaneously minimizing the cost of the defenses needed to lower the vulnerability. 
We employ two separate objective functions: local and global (Eq.~(\ref{eqn:local_function}) and (\ref{eqn:global_function}), respectively). The local objective function is applied to every defense-graph pair in 
the current defense set; the pair that minimizes the objective function is added to the defense 
history. The global objective function is employed \emph{after} the algorithm has completed populating 
the defense history; it chooses the optimal ``moment" (cumulative set of defenses) from the history. 
Each function employs a user-specified parameter $\alpha$ that tells the function how to weigh the 
adversary score against the cost. Note that in the local objective function, we weigh the marginal increase in defense cost against the corresponding marginal decrease in the graph's adversary score, while in the global objective function, we weigh the total cost against the current adversary score.

\subsubsection{Choosing Defenses}
\label{subsubsection:choosing_defenses}

\begin{algorithm}[h!]
\small

\SetKwFunction{FMain}{refresh\_Defense\_Set}
  \SetKwProg{Fn}{Function}{:}{}
  \Fn{\FMain{$d$, $G$}}{
    \KwData{Defense $d$ just selected by objective function, graph $G$ with optimal defense just added}

    \For{$s$ \emph{in} $S$}{ 

        \If{$s.timeInSet > max\_Set\_Time$}{
    
            $S$.remove($s$)\;
            $available\_Defense\_List$.add($s.defense$)
        
        }
    }
    
    \For{$G \in $ $S.graphs$} {
    
        $G$.apply\_Defense($chosen\_Defense$)\;
        $G$.propagate\_Defense($chosen\_Defense$)\;
    }
    
    $max\_Vul\_Device\_Defenses =$ All unused defenses from the device with the highest exploit risk score\;
    
    \While{ $\mid S \mid$ $< set\_Size$ \emph{and} $\mid available\_Defense\_List \mid$ $> 0$ }{
    
        \eIf{$\mid max\_Vul\_Device\_Defenses \mid$ $> 0$}{
        
            $d$ = random\_Defense($max\_Vul\_Device\_Defenses$)\;
        }{
            $d$ = random\_Defense($available\_Defense\_List$)\;
        }
        
        $new\_G =$ deep\_Copy($G$)\;
        $new\_G$.\textbf{apply\_Defense}($d$)\;
        $new\_G$.\textbf{propagate\_Defense}($d$)\CommentSty{; /* Similar to Section \ref{subsubsection:propagate_exploit_risk_scores}, except that the first nodes in the queue are the nodes upon which the new defense is placed */}\\
        $S$.add($d$, $new\_G$)\;
        $available\_Defense\_List$.remove($d$)\;
    }

    \KwRet{$S$}\;
  }
  
  \caption{Refreshing the defense set}
\label{fig:refresh_defense_set}
    
\end{algorithm}

In every optimization round, we choose the defense that minimizes the value of the local
optimization function: Eq.~(\ref{eqn:local_function}). The defenses that are used in this
comparison are located in the Defense set $S$, a list of fixed length that contains a subset of defenses that are most likely to improve the optimization function. Every defense in $S$ is associated with a graph $G$ that contains that same defense and all the defenses already applied to $G$. This list is managed via Algorithm \ref{fig:refresh_defense_set}, which is invoked to refresh the defense set after every optimization round. 

A defense may be removed from $S$ for two reasons: either it has already been selected as the optimal defense in the previous optimization round, or it has been in the set for too many optimization rounds (longer than $max\_Set\_Time$), which means that it probably does not contribute much to reducing the system's exploit risk. Defenses that are removed are added back to an $available\_Defense\_List$ (initially set to $M$), which contains all possible defenses that the optimization function can select minus those that have already been used. When a new defense is needed in $S$, the algorithm randomly selects a defense from the $available\_Defense\_List$ with a preference for defenses from the device that possesses the highest exploit risk score.

\subsection{Adversary Models}
\label{subsubsection:adversary_models}

The vulnerability of a given IoT system can be expressed using a function of its entry node exploit risk scores (see Eq.~(\ref{eqn:adversary_score})). This ``adversary score" contains information about the ease of exploitation and impact of exploit goals in the entire graph (see Section \ref{subsubsection:propagate_exploit_risk_scores} for more detail). Given a set of entry nodes $A$, an adversary would likely want to enter the system at its most vulnerable location(s). As a result, our optimization process should  minimize the exploit risk scores of the $k$ highest-scoring entry nodes.

\begin{equation}
\label{eqn:adversary_score}
adversary\_Score = \max\limits_{A^{\ast} \subset A \emph{, }\mid A^{\ast} \mid = k} \frac{\displaystyle\sum\limits_{i=1}^{k}a_i.exploit\_Score}{\large{k}}
\end{equation}

\par Using only the highest-scoring entry node is not recommended because the $adversary\_Score$ tends to ``plateau" (bottom out) after only a few defenses are added. This happens because there may be no additional defenses that substantially impact the highest-scoring nodes. For good performance (and to adequately protect all parts of a complex system), the system administrator should specify a $k$ that is at least equal to the total number of devices in order to include exploit risk in a variety of different locations. The administrator can also adjust the vulnerability of an entry node by changing its \textit{accessibility} score.

\par This approach assumes that the adversary has white-box knowledge about the system, and that we, the defenders, are knowledgeable enough about an adversary's motivations to confidently assign quantitative \textit{impact} scores to exploit goals. However, the adversary may not be fully knowledgeable about all the defenses that we have added to the system, and we may not fully understand the adversary's motivations. To account for this, the system administrator can instruct GRAVITAS to randomly select attack outcome \textit{impact} scores drawn from a distribution. By adding additional ``noise" to the model, the system administrator can ensure that their IoT/CPS is prepared to tackle a wide variety of adversaries and attacks.

\subsection{Parameter Validation}
\label{subsection:parameter_validation}

The vulnerability scoring process contains several additions and modifications from those used
in other models \cite{first2019specification, ur-rehman2019}. These include a unique ``activation function" for exploit risk calculation and several scoring factors used in the vulnerability node scoring process. In order to validate these changes, we created a system called TASC 
(Testing for Autonomously-Generated Smart Cities) that generates quasi-random IoT/CPS. Controllable 
parameters include the number of devices, the relative number of different device 
categories/subcategories, the distribution of connection types between different devices, defense 
types and costs, and \textit{impact} scores for exploit goals. In theory, large systems generated with the 
same parameters but with a different random seed should have similar properties and broadly similar 
optimization curves. When optimized using the same adversary model and propagation/optimization 
parameters (such as $\alpha\_Local$), these systems should trace a similar adversary score 
vs. defenses-added curve, and should also possess a similar globally-optimal solution for a given 
$\alpha\_Global$.

\par One of the novel features of GRAVITAS involves its defense-addition optimization process. Adding a defense changes the vulnerability score of the nodes and edges affected by it, which in turn influences the exploit risk scores at the entry nodes once re-propagation is completed. The adapted scoring factors in Table \ref{fig:CVSS_scores} were chosen by generating several broadly similar systems using different seeds and employing different scoring sets in each system's optimization procedure to identify which set resulted in the most consistent results. We chose the activation function for the union probability function (see Section \ref{subsubsection:propagate_exploit_risk_scores}) in a similar manner, comparing several different functions (including exponential, power, and logistic functions) with several different numeric values for each. By creating TASC systems with the same parameters but different random seeds, we were able to determine which defense score set and activation function combination was most likely to produce consistent results.

\section{Case Study: Smart Home}
\label{section:example_application}

\par
\begin{figure}[h!]
\centering
\includegraphics[scale=0.45]{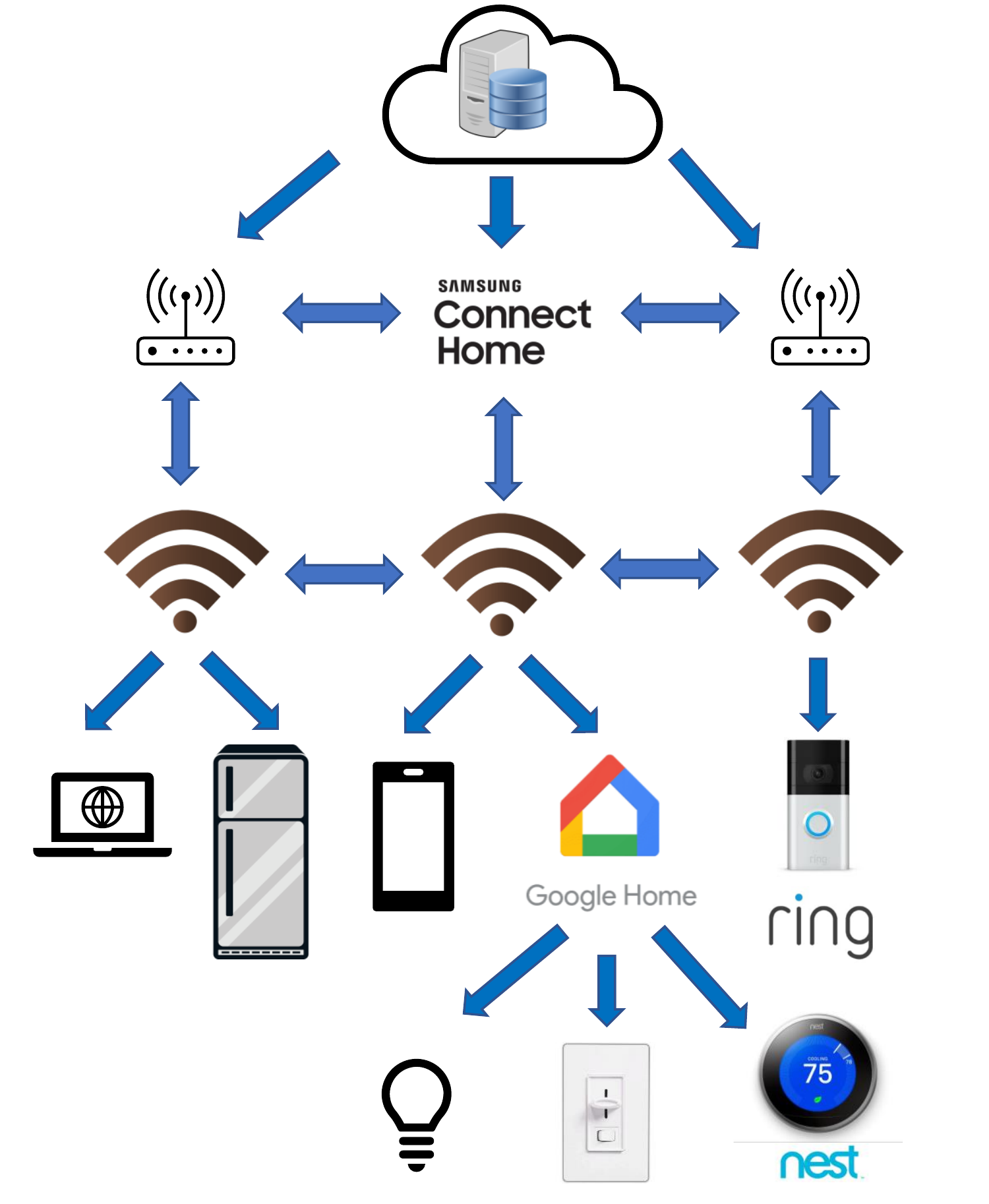}
\caption{A simplified representation of the smart home system. Our experimental setup contains 23
devices, including 3 WiFi routers, 4 local controllers, 11 sensors/actuators, and over 50 connections.} 
\label{fig:smart_home_system}
\end{figure}

 To demonstrate the functionality of GRAVITAS, we simulated a sample smart home system involving 
common household devices. Fig.~\ref{fig:smart_home_system} shows a simplified version of the system 
used for this analysis. While real-world devices in this system would contain built-in defenses 
out-of-the-box, this system assumes that the devices initially contain no defenses so that GRAVITAS 
can add them. Our defense set $M$ consists of the IoT-specific defenses outlined in ENISA's \textit{Baseline Security Recommendations for IoT}  \cite{enisa2017}. 

\begin{figure}[h!]
\centering
\includegraphics[scale=0.65]{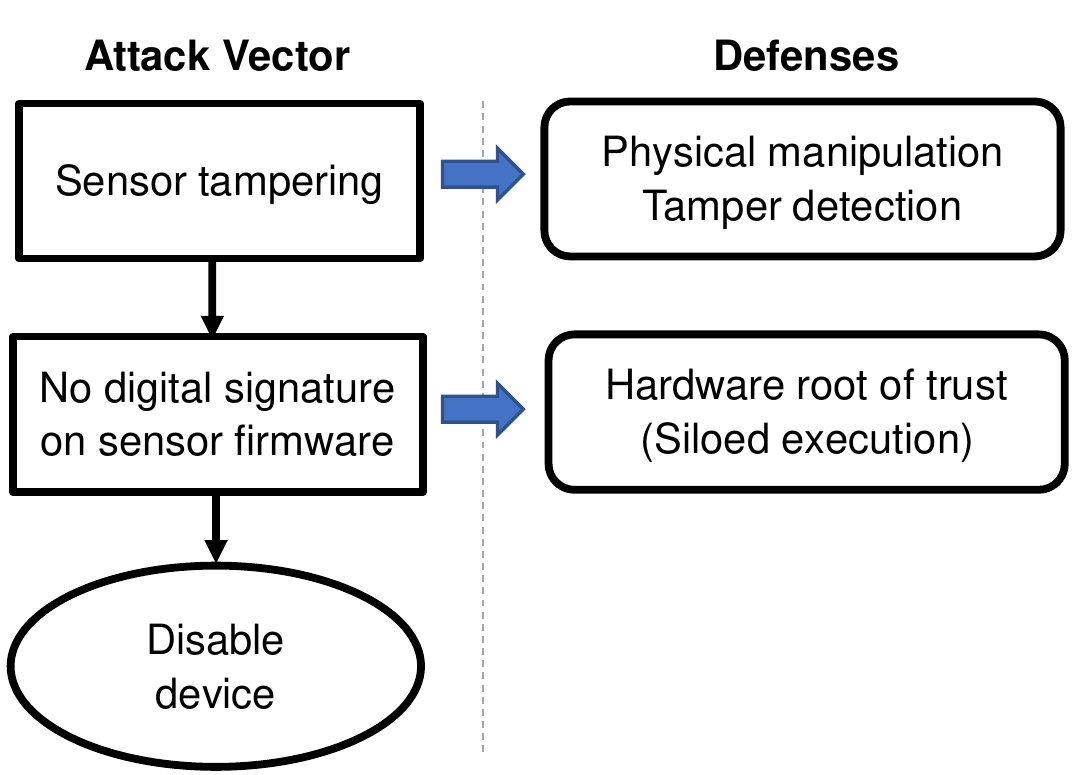}
\caption{An exploit that employs vulnerabilities in a single device: the Ring garden spotlight camera. The defenses selected by GRAVITAS are shown on the right.} 
\label{fig:single_exploit}
\end{figure}

\begin{figure}[h]
\centering
\includegraphics[scale=0.62]{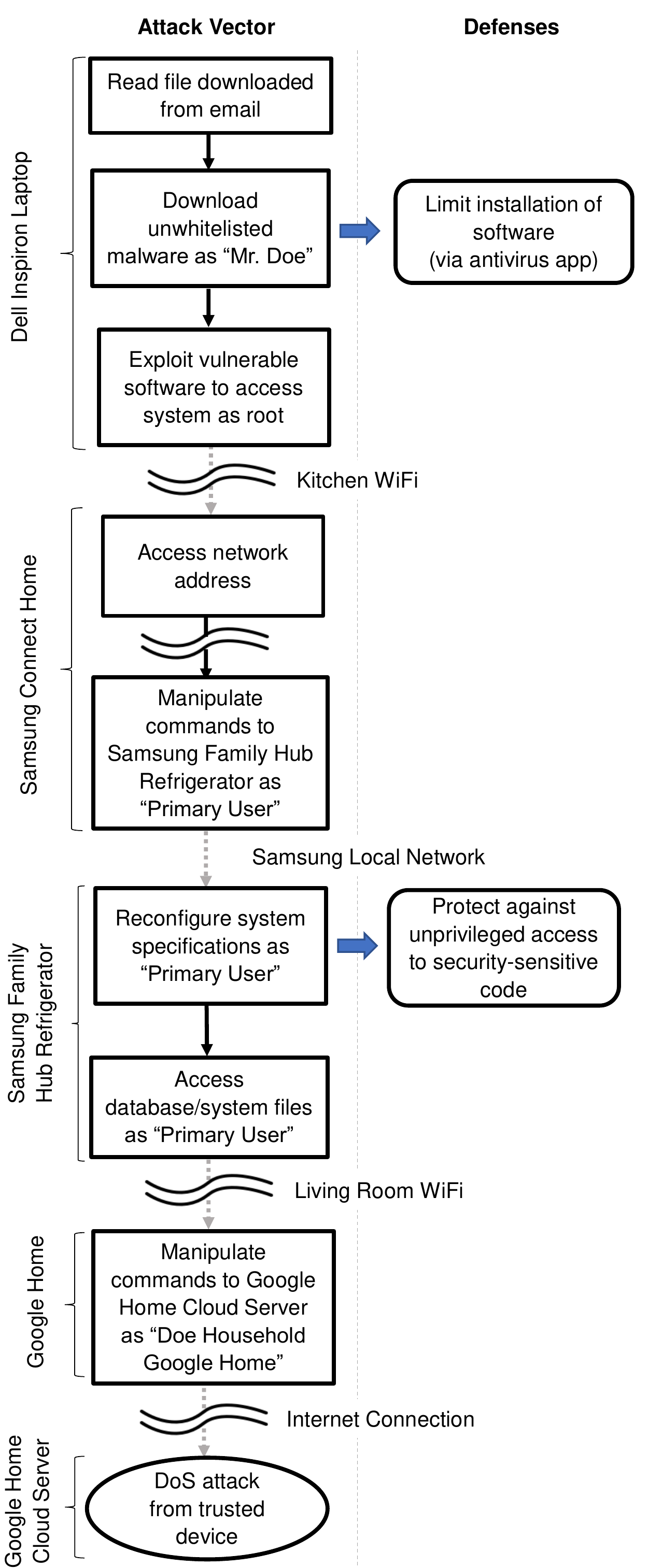}
\caption{An exploit that employs vulnerabilities in multiple devices. Since there are 17 vulnerabilities in the exploit, not all vulnerabilities (and defenses) in the exploit are shown.} 
\label{fig:multiple_exploit}
\end{figure}

\par GRAVITAS's novelty lies in its ability to discover attack vectors (exploit chains) passed over by traditional risk management software. Not only is it effective in discovering attack vectors in a single device (Fig. \ref{fig:single_exploit}), but it can also discover multi-stage attack vectors that consist of exploiting vulnerabilities in multiple different devices (Fig. \ref{fig:multiple_exploit}). These attacks might involve an adversary accessing trusted edge-side devices to gain access to private information in a cloud server, or using illicitly-obtained access to a central controller or user device to disable mission-critical sensing equipment. Both attacks demonstrate GRAVITAS's ability to include IoT/CPS-specific vulnerabilities in the attack graph, such as physical tampering for edge-side devices. There are numerous paths through the system available to the adversary, and GRAVITAS can find those attack vectors that are most likely to be targeted.

\begin{figure*}[t!]
\centering
\includegraphics[scale=0.91]{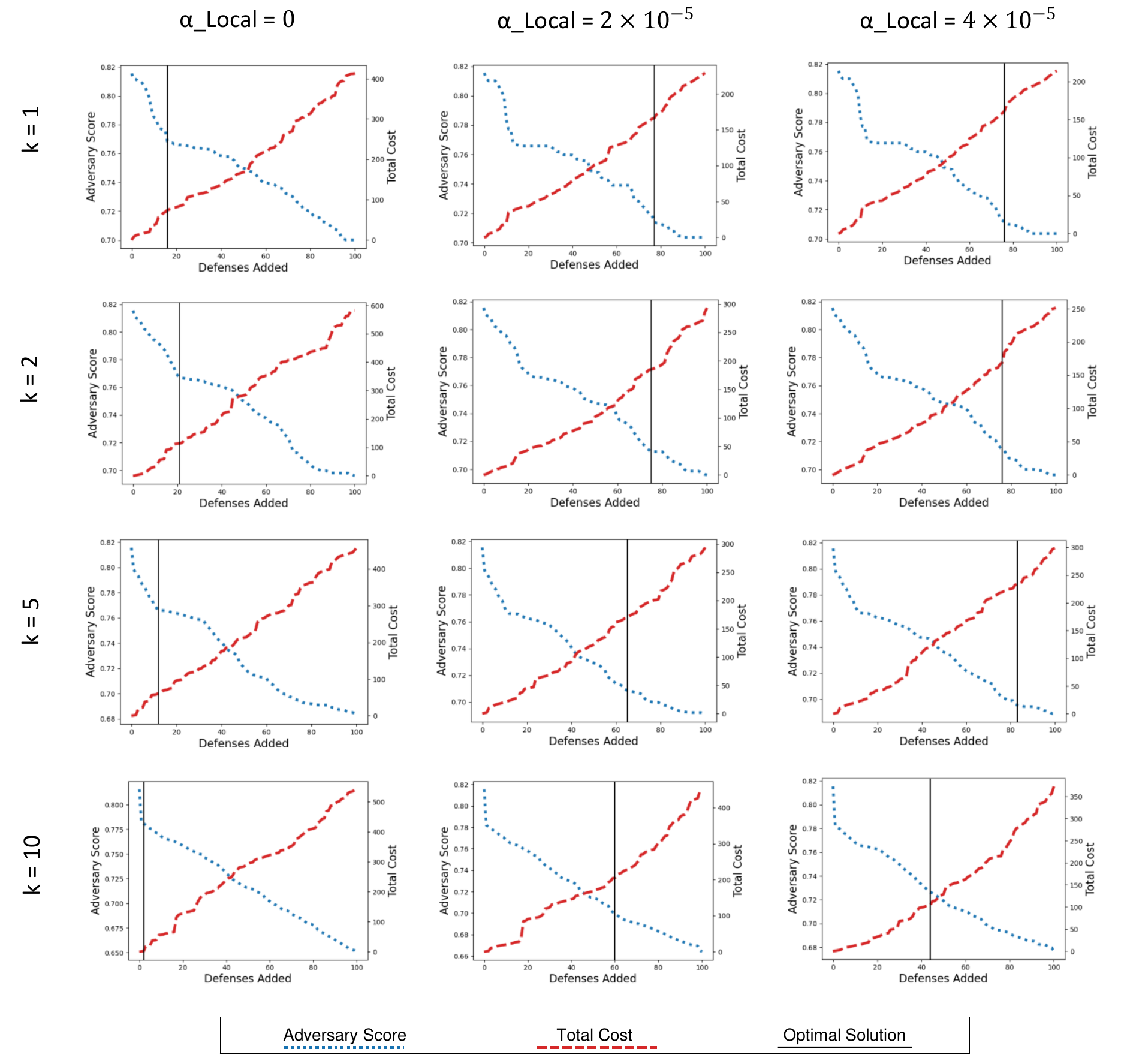}

\caption{Optimization curves and global optimal solutions during optimization of the 
smart home system. $\alpha\_Local$ represents the parameter for the local objective function 
(higher $\alpha\_Local$ means higher cost sensitivity), while $k$ represents the number of entry 
nodes whose exploit risk scores are averaged to obtain the adversary score. The black 
line represents the global optimal solution for $\alpha\_Global = 0.00032$.} 
\label{fig:example_graphs}
\end{figure*}

\par In Fig. \ref{fig:multiple_exploit}, we observe that the end result of the exploit is a DoS attack on the cloud server for the Google Home device. Its origin is a malware downloaded onto a laptop, which in turn instructs the Samsung fridge to flood the Google Home server with bogus traffic, thus launching a DoS attack. All this communication occurs via intermediaries such as a Samsung Connect controller and a WiFi router. Although the exploit looks complex, the adversary only needs to know that the laptop user has the Samsung Connect software installed. As a result of the length of this exploit chain, only a select few defenses are necessary to dramatically lower the exploit risk. 

\par The system administrator can view these exploits independently or use GRAVITAS to determine the vulnerabilities that constitute the ``weakest link" and disproportionately contribute to the adversary score. Table \ref{table:weakest_link} shows the three vulnerabilities with the highest exploit risk 
before defenses are added. Most traditional vulnerability scan-based risk management 
systems would not mark these vulnerabilities as particularly noteworthy since each cannot 
significantly compromise the security of its resident device by itself. However, GRAVITAS recognizes 
the importance of these vulnerabilities due to their crucial role in several attack vectors that lead 
to attack outcomes with high impact scores. It is worthwhile to also note that 
these exploit risk scores may be located in vulnerability nodes that are not entry 
nodes ($n \notin A$); as a result, the entry node exploit risk scores may be lower due to score 
dilution during the score propagation process.	
		
\begin{table}[h!]
    \centering
    \caption{The smart home vulnerabilities with the highest exploit risk score}
    \label{table:weakest_link}
    \begin{tabular}{|M{2.8cm}|c|c|}
        \hline
        \textbf{Device} & \textbf{Vulnerability} & \textbf{Exploit Risk Score}  \\ \hline
         Ring Base Station & Access requested & 0.7190 \\ \hline
         Ring Front Door Motion Sensor & Sensor tampering & 0.7040 \\ \hline
         Ring Garden Motion Sensor & Sensor tampering & 0.7035 \\ \hline
         
    \end{tabular}
\end{table}
		
\par This smart home system illustrates how apparently non-fatal vulnerabilities of stand-alone IoT devices can be combined to execute a fatal attack on a multi-device system. One example of this is the ``Reconfigure system specifications as primary user" vulnerability of the Samsung Family Hub Refrigerator in Fig. \ref{fig:multiple_exploit}. Generally speaking, gaining administrator-level access to an isolated smart fridge would not be a particularly harmful attack. However, because of the fridge’s connection to Google Home and its subsequent connection to the Google Home Cloud Server, that 
one vulnerability can be used to execute a potent DoS attack via the smart home network hub; this 
could result in Google blacklisting the hub's IP address, thus disabling the entire IoT system. 
This type of attack may not be significant for a small home-based IoT system, but for a Smart Factory, 
it may have significant economic and safety consequences.

\par Fig.~\ref{fig:example_graphs} shows the result of the optimization process using different values of $\alpha\_Local$ (the local objective function weighting parameter) and $k$ (the number of entry node exploit risk scores averaged to obtain the adversary score). These graphs in Fig.~\ref{fig:example_graphs} show that the adversary score curve generally 
becomes less ``noisy" as $k$ increases. This is expected given that we are averaging more scores for 
higher $k$. As $\alpha\_Local$ increases, cost is weighted higher in the objective function, and 
thus the adversary score drops less quickly while the cost also increases less quickly. Changing $k$ 
or the adversary score has a nonlinear impact on the optimal value (represented by the black line for $\alpha\_Global = 0.00032$).

\section{Discussion}
\label{section:discussion}

To enhance the current framework, it may be necessary to occasionally prune the graph and incorporate software-specific vulnerabilities from repositories like the NVD. Tools such as CVE-Search can be used to 
extract an up-to-date list of vulnerabilities for every device, while off-the-shelf software can be employed to match these vulnerabilities with those in the 
template attack graphs \cite{ghazo2019, dulaunoy2020, jajodia2009, openvas}. Yet even without these additions, GRAVITAS still provides added value in that it considers the whole array of vulnerabilities likely to be present in a device, rather than only those that have already been discovered. This approach allows proactive management of the system, allowing the system administrator to fortify the devices and network connections that would have the highest risk of exploitation if the vulnerability existed.

\par The categories and subcategories try to encompass the wide range of IoT/CPS devices that exist today, but they cannot realistically be expected to cover all devices. For example, the authentication procedures of embedded devices vary widely, not least because of the dozens of different security protocols and miniature operating systems that these devices use \cite{hahm2016}. In addition, the limited graph templates of GRAVITAS may lead the system administrator to ``pigeon-hole"  their device into an ill-fitting category, accidentally including vulnerabilities that may not exist and excluding those that do exist. Fortunately, GRAVITAS was built to be adaptable: system administrators can easily add additional device templates and known vulnerabilities to the system graphs. As a result, GRAVITAS is adaptable to applications in 5G systems, the design of medical body-area networks, smart city systems, manufacturing facilities, and public utility networks.

\par The optimization component of GRAVITAS could also be improved. Its ``greedy local search" 
methodology does not consider how adding a defense in the current round 
will affect the objective function value in later rounds. One possibility is to add ``lookahead" 
functionality that simply adds multiple defenses instead of one to each new entrant in the defense set. A 
more complex approach would see the local objective function augmented with information about 
past iterations of the attack graph, perhaps employing a nonlinear estimator for future defense 
additions.

\par Despite the cyclic nature of the attack graphs, there were never any convergence issues during 
the millions of exploit risk score propagation cycles we ran during parameter validation 
and the smart home tests. Virtually every propagation iteration completed in 50 cycles or less, 
and none exceeded 100 cycles of propagation. However, there were some issues regarding the consistency 
of vulnerability scores among different propagation cycles: the exact same graphs with the exact same 
defenses applied in a different order would sometimes have a slightly different maximum vulnerability 
score. This is likely due to rounding errors that compound after several thousand floating-point 
calculations during score propagation. While these errors were immaterial in the Smart 
Home system and the TASC systems, they do point to the possibility of a ``Butterfly Effect" phenomenon in larger systems, 
where a different random seed or slightly-modified system can have a larger impact on the 
outcome. To guard against this, the system administrator should perform the entire optimization process multiple times with different seeds so that there are several defense histories from which to pick the optimal solution.

\par Another issue with GRAVITAS lies in its treatment of defenses. Adding new hardware defenses is 
difficult post-production, and although it is theoretically possible to add software updates to an 
existing device, this is not always feasible. For an extant IoT/CPS, rearranging the connections 
between devices is often far more feasible than changing the devices themselves. Future versions of 
GRAVITAS should not just be able to add node and edge defenses, but also rearrange the system 
topology, including permissions and local network connections.

\section{Conclusion}
\label{section:conclusion}

GRAVITAS provides new insights into the security of complex IoT/CPS, suggesting new exploits by incorporating potential vulnerabilities into the attack graph and applying strategically-placed defenses that reduce the system's exploit risk. It is also adaptable to the risk model of the system administrator and can be used to determine which assets are most likely to be impacted in the event of a system breach. GRAVITAS optimizes the defense placements to obtain the best trade-off between system security and cost of operation. Most importantly, GRAVITAS allows for an organization to test and repair the design of an IoT/CPS before deployment, providing a proactive security 
solution that takes a holistic view of the system. In an era where IoT/CPS will soon be ubiquitous, 
getting the security right the first time is essential. As a risk assessment tool specifically tailored to 
the unique devices and complex topology of IoT/CPS, GRAVITAS could become an important tool in the 
arsenal of security practitioners.


\ifCLASSOPTIONcaptionsoff
  \newpage
\fi

\bibliographystyle{IEEEtran}
\bibliography{bibtex/bib/IEEEexample}

\begin{thebibliography}{10}
\providecommand{\url}[1]{#1}
\csname url@samestyle\endcsname
\providecommand{\newblock}{\relax}
\providecommand{\bibinfo}[2]{#2}
\providecommand{\BIBentrySTDinterwordspacing}{\spaceskip=0pt\relax}
\providecommand{\BIBentryALTinterwordstretchfactor}{4}
\providecommand{\BIBentryALTinterwordspacing}{\spaceskip=\fontdimen2\font plus
\BIBentryALTinterwordstretchfactor\fontdimen3\font minus
  \fontdimen4\font\relax}
\providecommand{\BIBforeignlanguage}[2]{{%
\expandafter\ifx\csname l@#1\endcsname\relax
\typeout{** WARNING: IEEEtran.bst: No hyphenation pattern has been}%
\typeout{** loaded for the language `#1'. Using the pattern for}%
\typeout{** the default language instead.}%
\else
\language=\csname l@#1\endcsname
\fi
#2}}
\providecommand{\BIBdecl}{\relax}
\BIBdecl

\bibitem{mosenia2017comprehensive}
A.~Mosenia and N.~K. Jha, ``A comprehensive study of security of
  {Internet-of-Things},'' \emph{IEEE Trans. Emerging Topics in Computing},
  vol.~5, no.~4, pp. 586--602, 2017.

\bibitem{idc2020}
\BIBentryALTinterwordspacing
{IDC}. (2020) {Worldwide Spending on the Internet of Things Will Slow in 2020
  Then Return to Double-Digit Growth}. Report. IDC Corporate USA. [Online].
  Available: \url{https://www.idc.com/getdoc.jsp?containerId=prUS46609320}
\BIBentrySTDinterwordspacing

\bibitem{bosche2018}
\BIBentryALTinterwordspacing
A.~Bosche, D.~Crawford, D.~Jackson, M.~Schallehn, and C.~Schorling. (2018)
  Unlocking opportunities in the {Internet of Things}. Report. Bain and
  Company. San Francisco, California. [Online]. Available:
  \url{https://www.bain.com/contentassets/5aa3a678438846289af59f62e62a3456/bain_brief_unlocking_opportunities_in_the_internet_of_things.pdf}
\BIBentrySTDinterwordspacing

\bibitem{akmandor2018smart}
A.~O. Akmandor and N.~K. Jha, ``Smart health care: An edge-side computing
  perspective,'' \emph{IEEE Consumer Electronics Magazine}, vol.~7, no.~1, pp.
  29--37, 2018.

\bibitem{stojkoska2017review}
B.~L.~R. Stojkoska and K.~V. Trivodaliev, ``A review of {Internet of Things}
  for smart home: Challenges and solutions,'' \emph{J. Cleaner Production},
  vol. 140, pp. 1454--1464, 2017.

\bibitem{zhang2014iot}
R.~Zhang and X.~Liu, ``{IoT}-based maintenance process design for fusion
  reactor remote handling system,'' \emph{J. Fusion Energy}, vol.~33, no.~6,
  pp. 653--657, 2014.

\bibitem{yun2010research}
M.~Yun and B.~Yuxin, ``Research on the architecture and key technology of
  {Internet of Things (IoT)} applied on smart grid,'' in \emph{Proc. IEEE Int.
  Conf. Advances in Energy Engineering}, 2010, pp. 69--72.

\bibitem{al2015role}
A.~Al-Ali and R.~Aburukba, ``Role of {Internet of Things} in the smart grid
  technology,'' \emph{J. Computer and Communications}, vol.~3, no.~05, p. 229,
  2015.

\bibitem{datta2016integrating}
S.~K. Datta, R.~P.~F. Da~Costa, J.~H{\"a}rri, and C.~Bonnet, ``Integrating
  connected vehicles in {Internet of Things} ecosystems: Challenges and
  solutions,'' in \emph{Proc. IEEE Int. Symp. A World of Wireless, Mobile and
  Multimedia Networks}, 2016, pp. 1--6.

\bibitem{thierer2015}
\BIBentryALTinterwordspacing
A.~Thierer and A.~Castillo. (2015) Projecting the growth and economic impact of
  the {Internet of Things}. Mercatus Center. Arlington, Virginia. [Online].
  Available: \url{https://www.mercatus.org/system/files/IoT-EP-v3.pdf}
\BIBentrySTDinterwordspacing

\bibitem{stavridis2016}
\BIBentryALTinterwordspacing
J.~Stavridis and D.~Weinstein. (2016, 11) The {Internet of Things} is a
  cyberwar nightmare. Foreign Policy Magazine. [Online]. Available:
  \url{https://foreignpolicy.com/2016/11/03/the-internet-of-things-is-a-cyber-war-nightmare/}
\BIBentrySTDinterwordspacing

\bibitem{lewis2016}
\BIBentryALTinterwordspacing
J.~A. Lewis. (2016, 2) Managing risk for the {Internet of Things}. Center for
  Strategic \& International Studies. [Online]. Available:
  \url{https://csis-website-prod.s3.amazonaws.com/s3fs-public/legacy_files/files/publication/160217_Lewis_ManagingRiskIoT_Web_Redated.pdf}
\BIBentrySTDinterwordspacing

\bibitem{markey2020}
\BIBentryALTinterwordspacing
E.~J. Markey and R.~Blumenthal. (2020, 6) Letter to acting administrator {James
  Owen} concerning cybersecurity issues with {Internet-Connected Cars}. United
  States Senate. [Online]. Available:
  \url{https://www.markey.senate.gov/imo/media/doc/NHTSA%20Cybersecurity%20Followup.pdf}
\BIBentrySTDinterwordspacing

\bibitem{margolis2017}
J.~{Margolis}, T.~T. {Oh}, S.~{Jadhav}, Y.~H. {Kim}, and J.~N. {Kim}, ``An
  in-depth analysis of the {Mirai} botnet,'' in \emph{Proc. Int. Conf. Software
  Security and Assurance}, 2017, pp. 6--12.

\bibitem{sehwag2016tv}
V.~Sehwag and T.~Saha, ``{TV-PUF: A} fast lightweight analog physical
  unclonable function,'' in \emph{Proc. IEEE Int. Symp. Nanoelectronic and
  Information Systems}, 2016, pp. 182--186.

\bibitem{mckay2016report}
K.~McKay, L.~Bassham, M.~S{\"o}nmez~Turan, and N.~Mouha, ``Report on
  lightweight cryptography,'' National Institute of Standards and Technology,
  Tech. Rep., 2016.

\bibitem{saha2020}
T.~{Saha}, N.~{Aaraj}, N.~{Ajjarapu}, and N.~K. {Jha}, ``{SHARKS}: Smart
  hacking approaches for risk scanning in {Internet-of-Things} and
  cyber-physical systems based on machine learning,'' \emph{IEEE Trans.
  Emerging Topics in Computing}, pp. 1--1, 2021.

\bibitem{nia2016}
A.~{Mosenia}, S.~{Sur-Kolay}, A.~{Raghunathan}, and N.~K. {Jha},
  ``Physiological information leakage: A new frontier in health information
  security,'' \emph{IEEE Trans. Emerging Topics in Computing}, vol.~4, no.~3,
  pp. 321--334, 2016.

\bibitem{matrosov2011}
\BIBentryALTinterwordspacing
A.~Matrosov, E.~Rodionov, D.~Harley, and J.~Malcho. (2011) Stuxnet under the
  microscope. ESET LLC. [Online]. Available:
  \url{https://www.esetnod32.ru/company/viruslab/analytics/doc/Stuxnet_Under_the_Microscope.pdf}
\BIBentrySTDinterwordspacing

\bibitem{hernandez2014}
\BIBentryALTinterwordspacing
G.~Hernandez, O.~Arias, D.~Buentello, and Y.~Jin. (2014) A smart {Nest}
  thermostat: A spy in your home. Black Hat USA. [Online]. Available:
  \url{https://www.blackhat.com/docs/us-14/materials/us-14-Jin-Smart-Nest-Thermostat-A-Smart-Spy-In-Your-Home.pdf}
\BIBentrySTDinterwordspacing

\bibitem{juels2003}
A.~Juels, R.~L. Rivest, and M.~Szydlo, ``The blocker tag: Selective blocking of
  {RFID} tags for consumer privacy,'' in \emph{Proc. ACM Conf. Computer and
  Communications Security}, 2003, pp. 103--111.

\bibitem{huang2011}
L.~Huang, A.~D. Joseph, B.~Nelson, B.~I. Rubinstein, and J.~D. Tygar,
  ``Adversarial machine learning,'' in \emph{Proc. ACM Wkshp. Security and
  Artificial Intelligence}, 2011, pp. 43--58.

\bibitem{ibm2019}
\BIBentryALTinterwordspacing
{IBM Security}. (2019) Penetration testing: Protect critical assets using an
  attacker’s mindset. IBM Corporation. [Online]. Available:
  \url{https://www.ibm.com/security/services/penetration-testing}
\BIBentrySTDinterwordspacing

\bibitem{ur-rehman2019}
A.~{Ur-Rehman}, I.~{Gondal}, J.~{Kamruzzuman}, and A.~{Jolfaei},
  ``Vulnerability modelling for hybrid {IT} systems,'' in \emph{Proc. IEEE Int.
  Conf. Industrial Technology}, 2019, pp. 1186--1191.

\bibitem{ghazo2019}
A.~T. {Al Ghazo}, M.~{Ibrahim}, H.~{Ren}, and R.~{Kumar}, ``{A2G2V}: Automatic
  attack graph generation and visualization and its applications to computer
  and {SCADA} networks,'' \emph{IEEE Trans. Systems, Man, and Cybernetics:
  Systems}, pp. 1--11, 2019.

\bibitem{tenable2021}
\BIBentryALTinterwordspacing
{Tenable Inc.} (2021) Tenable home page. Tenable Inc. [Online]. Available:
  \url{https://www.tenable.com/}
\BIBentrySTDinterwordspacing

\bibitem{ou2005}
X.~Ou, S.~Govindavajhala, and A.~W. Appel, ``{MulVAL}: A logic-based network
  security analyzer.'' in \emph{Proc. USENIX Security Symp.}, vol.~8, 2005, pp.
  113--128.

\bibitem{malowidzki2019}
M.~{Malowidzki}, D.~{Hermanowski}, and P.~{Berezinkki}, ``{TAG}: Topological
  attack graph analysis tool,'' in \emph{Proc. Cyber Security in Networking
  Conf.}, 2019, pp. 158--160.

\bibitem{jajodia2009}
S.~Jajodia and S.~Noel, \emph{Topological Vulnerability Analysis}.\hskip 1em
  plus 0.5em minus 0.4em\relax Center for Secure Information Systems, George
  Mason University, 2009, pp. 139--154.

\bibitem{lippmann2006}
R.~{Lippmann}, K.~{Ingols}, C.~{Scott}, K.~{Piwowarski}, K.~{Kratkiewicz},
  M.~{Artz}, and R.~{Cunningham}, ``Validating and restoring defense in depth
  using attack graphs,'' in \emph{Proc. IEEE Military Communications Conf.},
  2006, pp. 1--10.

\bibitem{first2019specification}
\BIBentryALTinterwordspacing
FIRST. (2019) Common vulnerability scoring system version 3.1 specification
  document. FIRST Inc. [Online]. Available:
  \url{https://www.first.org/cvss/v3-1/cvss-v31-specification_r1.pdf}
\BIBentrySTDinterwordspacing

\bibitem{enisa2017}
\BIBentryALTinterwordspacing
{The European Union Agency for Network and Information Security (ENISA)}.
  (2017, 11) {Baseline Security Recommendations for IoT} in the context of
  critical information infrastructures. [Online]. Available:
  \url{https://www.enisa.europa.eu/publications/baseline-security-recommendations-for-iot}
\BIBentrySTDinterwordspacing

\bibitem{mitre2021}
\BIBentryALTinterwordspacing
{The MITRE Corporation}. (2021) {MITRE Attack Mitigations - Enterprise}.
  [Online]. Available: \url{https://attack.mitre.org/mitigations/enterprise/}
\BIBentrySTDinterwordspacing

\bibitem{kordy2014}
B.~Kordy, L.~Piètre-Cambacédès, and P.~Schweitzer, ``{DAG}-based attack and
  defense modeling: Don’t miss the forest for the attack trees,''
  \emph{Computer Science Review}, vol. 13-14, pp. 1 -- 38, 2014.

\bibitem{ou2012}
X.~{Ou} and A.~{Singhal}, \emph{Quantiative Security Risk Assessment of
  Enterprise Networks}.\hskip 1em plus 0.5em minus 0.4em\relax New York, NY:
  Springer, 2012.

\bibitem{noel2010}
S.~Noel, L.~Wang, A.~Singhal, and S.~Jajodia, ``Measuring security risk of
  networks using attack graphs,'' \emph{Int. J. Next Generation Computing},
  vol.~1, Jan. 2010.

\bibitem{aksu2017}
M.~U. Aksu, M.~H. Dilek, E.~{\.I}. Tatl{\i}, K.~Bicakci, H.~I. Dirik, M.~U.
  Demirezen, and T.~Ayk{\i}r, ``A quantitative {CVSS}-based cyber security risk
  assessment methodology for {IT} systems,'' pp. 1--8, 2017.

\bibitem{first2019userguide}
\BIBentryALTinterwordspacing
FIRST. (2019) Common vulnerability scoring system version 3.1 user guide. FIRST
  Inc. [Online]. Available:
  \url{https://www.first.org/cvss/v3-1/cvss-v31-user-guide_r1.pdf}
\BIBentrySTDinterwordspacing

\bibitem{dulaunoy2020}
\BIBentryALTinterwordspacing
A.~Dulaunoy, P.~Moreels, and R.~Vinot. (2020) {CVE}-search project. CVE-Search.
  [Online]. Available: \url{http://cve-search.org}
\BIBentrySTDinterwordspacing

\bibitem{openvas}
\BIBentryALTinterwordspacing
{OpenVAS - Open Vulnerability Assessment Scanner}. Greenbone Networks GmbH.
  [Online]. Available: \url{https://www.openvas.org/}
\BIBentrySTDinterwordspacing

\bibitem{hahm2016}
O.~{Hahm}, E.~{Baccelli}, H.~{Petersen}, and N.~{Tsiftes}, ``Operating systems
  for low-end devices in the {Internet of Things}: A survey,'' \emph{IEEE
  Internet of Things Journal}, vol.~3, no.~5, pp. 720--734, 2016.

\end{thebibliography}

\newpage
\begin{IEEEbiography}[{\includegraphics[width=1in,height=1.25in,clip,keepaspectratio]{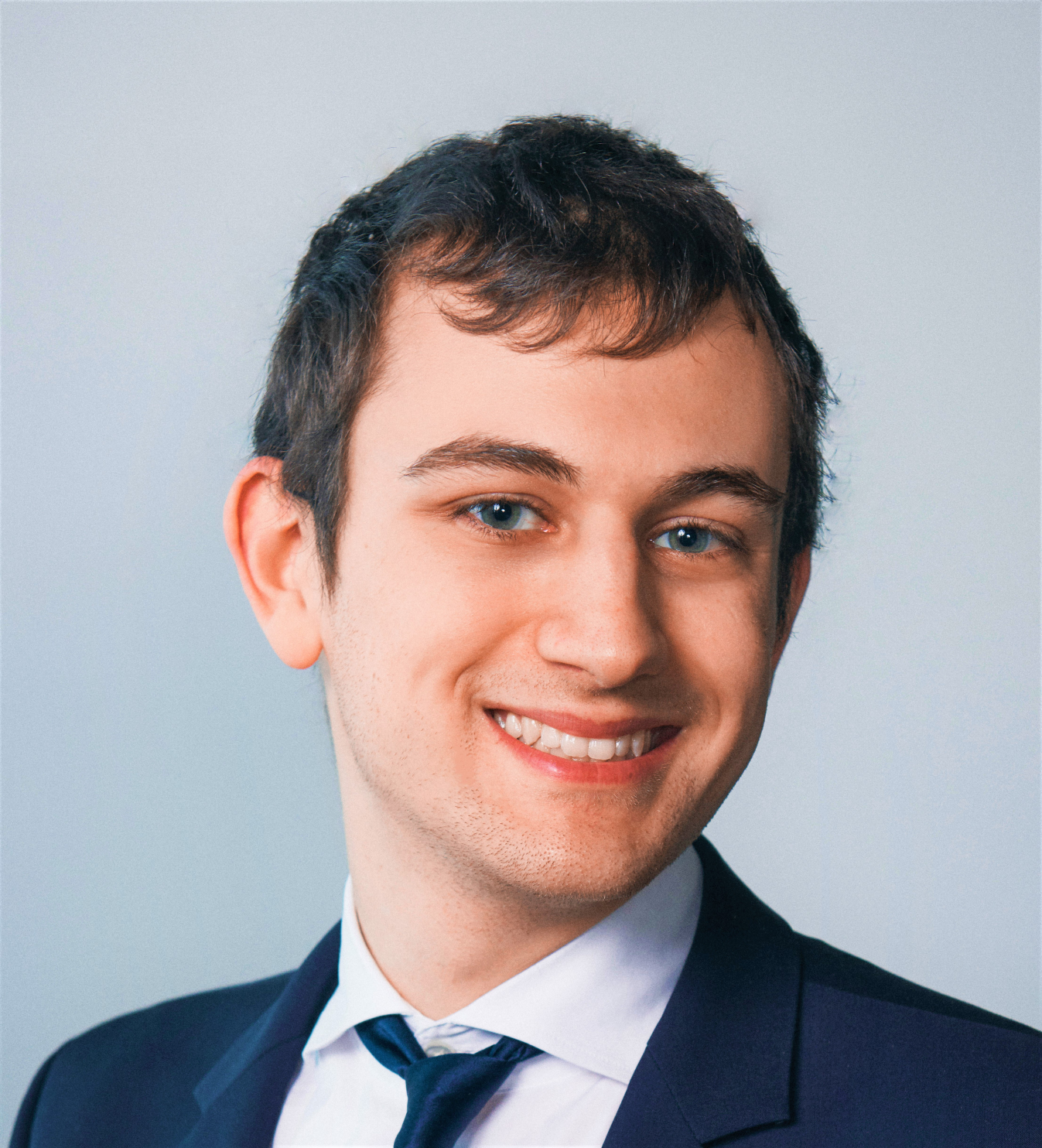}}]{Jacob Brown}
Jacob Brown received his Bachelor of Science in Electrical Engineering (Hons.) from Princeton 
University. He has conducted machine learning (ML) research for security and healthcare applications, 
such as creating a COVID-19 outbreak prediction model based on micro-level demographic data. Jacob 
has also worked on ML-based geospatial modeling of conflict zones. A member of Sigma Xi, he has 
received several grants and prizes in recognition of his work, including a grant from the Project X 
Innovation Fund and the G. David Forney Jr. Prize for an outstanding record in communication 
sciences.
\end{IEEEbiography}

\begin{IEEEbiography}[{\includegraphics[width=1in,height=1.25in,clip,keepaspectratio]{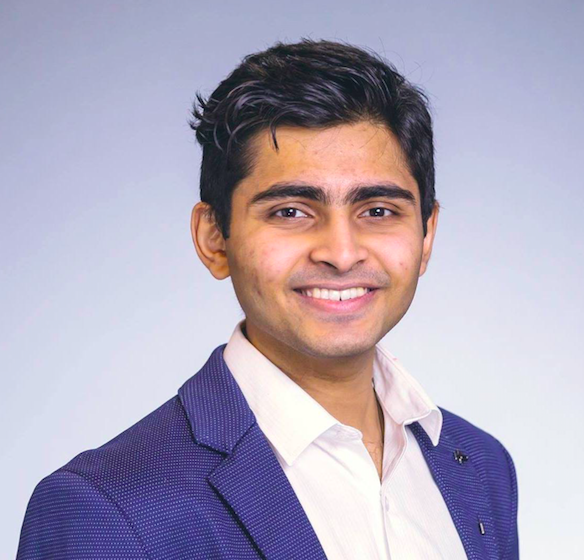}}]{Tanujay Saha}
Tanujay Saha is currently pursuing his Ph.D. degree in Electrical and Computer Engineering (ECE) at Princeton University, 
NJ, USA. He received his Bachelors in Technology (Hons.) from Indian Institute of Technology, 
Kharagpur, India in 2017. He has held research positions in various 
organizations and institutes like Intel Corp. (USA), KU Leuven (Belgium), and Indian 
Statistical Institute. His research interests lie at the intersection of 
IoT, cybersecurity, machine learning, embedded systems, and cryptography. He 
has extensive industry and academic experience in both theoretical and practical 
aspects of cryptography and machine learning.
\end{IEEEbiography}


\begin{IEEEbiography}[{\includegraphics[width=1in,height=1.25in,clip,keepaspectratio]{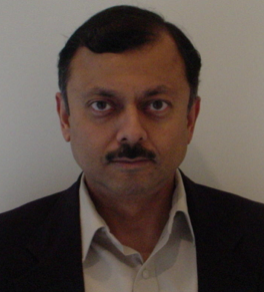}}]{Niraj K. Jha}
Niraj K. Jha received the B.Tech. degree in electronics and electrical
communication engineering from I.I.T., Kharagpur, India, in 1981, and the
Ph.D. degree in electrical engineering from the University of Illinois at
Urbana-Champaign, Illinois, in 1985. He has been a faculty member of the
Department of Electrical Engineering, Princeton University, since 1987.
He was given the Distinguished Alumnus Award by I.I.T., Kharagpur. He
has also received the Princeton Graduate Mentoring Award. He has served
as the editor-in-chief of the IEEE Transactions on VLSI Systems and as
an associate editor of several other journals. He has co-authored five
books that are widely used. His research has won 20 best paper awards or
nominations. His research interests include smart healthcare,
cybersecurity, machine learning, and monolithic 3D IC design. He has
given several keynote speeches in the area of nanoelectronic design/test
and smart healthcare. He is a fellow of the IEEE and ACM.
\end{IEEEbiography}




\end{document}